# Temporal Events Detector for Pregnancy Care (TED-PC): A Rule-based Algorithm to Infer Gestational Age and Delivery Date from Electronic Health Records of Pregnant Women with and without COVID-19

**Tianchu Lyu[1], Chen Liang[1], Jihong Liu[2], Berry Campbell[3], Peiyin Hung[1], Yi-Wen Shih[1], Nadia Ghumman[1], Xiaoming Li[4], on behalf of the National COVID Cohort Collaborative Consortium**

[1] Department of Health Services Policy and Management, Arnold School of Public Health, University of South Carolina
[2] Department of Epidemiology & Biostatistics, Arnold School of Public Health, University of South Carolina
[3] Department of Obstetrics and Gynecology, School of Medicine, University of South Carolina
[4] Department of Health Promotion Education and Behaviors, Arnold School of Public Health, University of South Carolina

Correspondence: Chen Liang, PhD, Assistant Professor, cliang@mailbox.sc.edu

**Abstract**

Objective: To develop a rule-based algorithm that detects temporal information of clinical events during pregnancy care for women with COVID-19 by inferring gestational age in weeks and delivery dates from Electronic Health Records (EHR) from the National COVID Cohort Collaborate (N3C).

Materials and Methods: The EHR are normalized by the Observational Medical Outcomes Partnership (OMOP) Clinical Data Model (CDM). EHR phenotyping resulted in 270,897 pregnant women (2018-06-01 to 2021-05-31). We developed a rule-based algorithm and performed a multi-level evaluation to test the content validity and clinical validity of the algorithm; and extreme value analysis for testing the algorithm for individuals with <150 or >300 days of gestation.

Results: The algorithm identified 296,194 pregnancies (16,659 COVID-19 174 and 744 without COVID-19 peri-pandemic) in 270,897 pregnant women. For inferring gestational age, 95% of identified cases (n=40) have moderate-high accuracy (overall Cohen's Kappa = 0.62); 100% of identified cases (n=40) have moderate-high granularity of temporal information (overall Cohen's Kappa = 1). For inferring delivery dates, the accuracy is 100% (overall Cohen's Kappa = 1). Accuracy of gestational age detection for extreme length of gestation is 93.3% (overall Cohen's Kappa = 1). There were 104,791 and 191,403 gestations before and during the COVID-19 pandemic. Compared with those mothers who were not infected, mothers with COVID-19 record showed higher prevalence in multiple medical conditions such as obesity (35.1% vs. 29.5%), diabetes (17.8% vs. 17.0%), chronic obstructive pulmonary disease (COPD) (0.2% vs. 0.1%), respiratory distress syndrome (ARDS) (1.8% vs. 0.2%).

Discussion: We explored the characteristics of pregnant women by different timing of COVID-19 with our newly developed algorithm: the first algorithm to infer temporal information from complete antenatal care and detect the timing of SARS-CoV-2 infection for pregnant women using N3C.

Conclusion: The algorithm shows excellent validity in inferring gestational age and delivery dates, which supports national EHR cohorts on N3C studying the impact of COVID-19 on pregnancy.

**Keywords: Electronic Health Records, information extraction, pregnancy, algorithm, COVID-19**

## Introduction

Recent findings suggested Coronavirus Disease 2019 (COVID-19) to be associated with an increased risk for adverse pregnancy outcomes and neonatal complications.[1,2] However, there has been limited knowledge pertaining to the timing of SARS-CoV-2 infection during the pregnancy (e.g., infection in a specific trimester, gestational age, or labor and delivery) and its association with pregnant women's real-time clinical presentation.[3] Timing of viral infection is important because fetuses are more vulnerable to maternal complications and/or viral infection during certain gestations.[4,5] Nevertheless, detecting the timing of viral infection poses substantial challenges to the quality of data and clinical information extraction methodology.

Pregnancy care consists of antenatal, labor and delivery, and postpartum care.[6] Because pregnancy care spans up to 21 months, it involves exceedingly rich clinical data. A complete episode of pregnancy care often involves multiple encounters with multiple health providers, clinical sites, and diverse clinical information systems, meaning that a vast number of clinical events are generated at an unprecedented granularity and data quality. Over decades, comprehensive clinical data for pregnancy care have not been widely available until recent advances in the use of normalized multi-system electronic health records (EHR), such as the National COVID Cohort Collaborative (N3C)[23], which provide growing Real-World Evidence (RWE) to support pregnancy research during COVID-19 pandemic.[7-10]

One of the unique characteristics of EHR in pregnancy care is the complex temporal relations of clinical events. To better understand the impacts of COVID-19 infections or COVID-19 pandemic on pregnancy health, it is important to know the length of pregnancy and the timing of pregnancy-related complications or events in relation to the duration of pregnancy (i.e., gestational age). Both mothers and fetuses experience crucial physiological changes and clinical complications during the pregnancy, which generates a substantial number of clinical events in the EHR. Accurate identification of temporal relations of clinical events across the entire episode of pregnancy care is a fundamental step for clinical decision making as well as downstream EHR data mining. Gestational age is also the prerequisite for tracing the timing of SARS-CoV-2 infection and COVID-19 vaccination, health care utilization, and causal pathway to adverse clinical outcomes. Although EHR data have the unique advantages in preserving temporal information of these clinical events, clinical information extraction methods tailored for pregnancy care have been scarce.[12]

Challenges in such clinical information extraction are manifold. First, controlled vocabularies (e.g., ICD, SNOMED-CT, LOINC, RxNorm) along with the relational EHR database architecture are designed to preserve some temporal information of clinical events, yet the information suffers from a low level of granularity (e.g., LOINC 95656-5: Gestational age [GA] <30 weeks, LOINC 49085-4: First and Second trimester integrated maternal screen panel), unreliable data entries (e.g., laboratory test results sometimes have delayed time stamps), and incomplete data (e.g., laboratory test results sometime are missing due to many laboratory results are photocopies). Second, approximately 80% of the EHR data consist of unstructured data (i.e., clinical notes), with which a considerable amount of temporal information is in the form of free text that cannot be directly used for quantitative analysis.[13] Third, EHR data for pregnant women are distinct from most other medical specialties in that pregnancy care has scheduled routine visits that are involved with antenatal care, labor and delivery hospitalization, and postpartum care. Incomplete and/or inconsistent data are common because patients are often engaged with different providers, health care systems, and non-pregnancy clinical visits with important data relevant to pregnancy.[12] For example, antenatal care and other medical care during the pregnancy could be at locations different from labor and delivery hospitals. Missing critical information from one clinical site would require researchers to infer such information using data from other sites or other visits. Health records of individual visits span inpatient and various outpatient visits but may not contain explicit and consistent temporal information (e.g., last menstrual period [LMP] [11], estimated date of delivery [DOD], and GA).

Current informatics methods for extracting and inferring temporal relations of clinical events include rule-based methods, machine learning, natural language processing (NLP), ontology-based methods, and temporal reasoning.[13-21] Most of these methods utilized unstructured clinical notes in combination with structured EHR data, which is comprehensive for generic temporal information extraction. However, informatics methods focusing on temporal events among pregnant women's EHR are limited. Among studies that extract or infer DOD and GA, LMP and imaging/lab test results are commonly used data; chart review is a commonly used method.[15,18,20,21] Using LMP data requires the providers to accurately document LMP in the EHR, yet in the real world, many EHR datasets have a lot of missing values in LMP. The use of ultrasound test results or other laboratory test results requires the EHR to comprehensively document both laboratory orders and testing results, yet testing results are often missing in real-world EHR datasets. Additionally, using laboratory data alone for inferring GA may not be accurate due to the individual physiological variation among pregnant women. A recent study utilized comprehensive ICD codes of diagnoses and procedures to infer delivery dates.[22] While the study focused on full-term pregnancy with comprehensive medical records during the labor and delivery hospitalization, comprehensive methods remain needed for early-stage pregnancy (e.g., extreme preterm, very preterm) and those who have part of the pregnancy care data and conflicting data documented in EHR. Particularly, there is no published methods for extracting temporal relations of clinical events for pregnant women with COVID-19, in which temporal relations of clinical events suggestive for exact time of viral infection, acute phase of COVID-19, and vaccinations are unique yet challenging for studying pregnant women with COVID-19.

To identify temporal relations of clinical events imperative for pregnant women with COVID-19, we developed a rule-based algorithm, namely Temporal Events Extractor for Pregnancy Care (TED-PC), that can infer GA and DOD using both structured EHR data and annotated clinical notes. The algorithm is designed to capture any temporal information to be used for inferring GA and DOD so that the complete temporal relations in a pregnancy episode can be replicated and the timing of SARS-CoV-2 infection (in weeks) can be detected. This design is anticipated to be effective for pregnant women with regular labor and delivery hospitalization, without complete hospitalization records, and those who have pre-term delivery, miscarriage, early-stage pregnancy and termination, and multiple births. This algorithm is specialized for EHR from the National COVID Cohort Collaborate (N3C) because N3C has 1) individual-level data linked among multiple health systems nationwide and 2) normalized procedures, laboratory tests/results, and annotated clinical notes, which enable the reasoning of GA and DOD for patients who commonly have missing and conflicting data. These unique characteristics of N3C are critical for detecting the temporal information for pregnant women with COVID-19. The performance and clinical validity of TED-PC are tested systematically on the N3C platform. Presently, this algorithm is used as a critical clinical information extraction tool to identify comprehensive temporal relations of clinical events from multiple COVID-19 pregnant women cohorts on N3C. It provides a high-throughput informatics solution to the urgent need for mining large-scale pregnant women's EHR data for combating COVID-19.

## Materials and Methods

### *Data sources*

We used the N3C database, a multi-center clinical data repository that contains de-identified EHR data of individuals with COVID-19 blended with controls (i.e., non-COVID-19).[23] N3C currently has EHR and medical claims data from more than 73 healthcare systems and institutes across 50 states. The EHR data are normalized using the Observational Medical Outcomes Partnership (OMOP) Clinical Data Model (CDM).[23,24] In order to find the full clinical course of each pregnancy, the study cohort included women who met the following conditions: (1) have at least one childbirth between 2018-06-01 and 2021-05-31, (2) be aged between 15 to 49 years old at the DOD, and (3) have at least one GA-related record during the pregnancy.

Because the N3C database is normalized by OMOP CDM, we utilized the following resources for EHR phenotyping. The ATHENA vocabulary is used for retrieving OMOP CDM concept IDs and phenotyping patients with the GA and childbirth-related records. The Algorithms section details the design and the procedures for using the ATHENA vocabulary.

### *Algorithms*

To retrieve the full spectrum of each gestation in the EHR data among pregnant women, it is crucial to identify the start date (i.e., pregnancy start) and the end date (i.e., childbirth delivery date) of the pregnancy. The start date can be estimated by the GA-related records that indicate the GA (e.g., in weeks, in a range of weeks, or a particular trimester) and the date of the record. The end date can be estimated by the identification of the DOD. Because some pregnant women's EHR data only have either GA-related records or childbirth delivery records, we first estimated GA and DOD, respectively, which resulted in a cohort of pregnant women with estimated GA, denoted as the GA cohort, and a cohort of pregnant women with estimated DOD, denoted as the DOD cohort. Then we estimated the start date and end date of the pregnancy by consolidating the temporal information from both cohorts. (Figure 1).

#### Gestational age cohort (GA cohort)

*Phenotyping.* The purpose of this step is to find OMOP CDM concepts that can be used to retrieve GA-related information from EHR data. For phenotyping the GA cohort, we used a keyword search strategy followed by a review of retrieved OMOP CDM concepts. Using the ATHENA vocabulary, a set of keywords were reviewed and determined: "trimester", "gestation", and "pregnan" (regular expression of "pregnancy" or "pregnant"). These keywords were used in conjunction with three filters in the ATHENA database: (1) "DOMAIN", which include "Condition", "Observation", "Procedure", "Measurement", etc.; (2) "CONCEPT", which include "Standard" and "Non-standard"; (3) "VALIDITY", which include "Valid" and "Invalid". The pseudo-query is:

*("trimester" OR "gestation" OR "pregnan") AND (((DOMAIN = "Condition") OR (DOMAIN = "Observation") OR (DOMAIN = "Procedure") OR (DOMAIN = "Measurement")) AND (CONCEPT = "Standard") AND (VALIDITY = "Valid"))*

In the review of the returned OMOP CDM concepts, we applied the following three criteria to narrow down the scope step by step to our target concepts: (a) "Whether a record indicates a pregnant patient"; (b) "If yes, whether the record contains GA information of the patient"; (c) "If yes, what is the value of the GA". Finally, we identified 138 OMOP CDM concepts (See Appendix 1). The researcher (TL) who performed the phenotyping was not involved in the phenotyping evaluation.

*Rule-based algorithm.* We developed a rule-based algorithm to infer GA from EHR data (Figure 2). A critical feature of the algorithm is that we divided all extracted OMOP CDM concepts into four accuracy levels based on their clinical meanings and granularity of the date: high, moderate-high, moderate-low, and low (Table 1) and prioritized the retrieval of GA-related information based on accuracy levels. Table 2 shows the pseudocode for the algorithm.

*Childbirth delivery cohort (DOD cohort)*

*Phenotyping.* For phenotyping the DOD cohort, we started with a list of CDC-recommended ICD, DRG, and CPT codes used for childbirth delivery and followed by exploring the relevant OMOP CDM concepts using the semantic relationships of concepts on the ATHENA vocabulary. First, we used a set of ICD-10, DRG, and CPT codes suggestive of childbirth delivery (see Appendix 2).[25] These codes were used to retrieve corresponding OMOP CDM concepts in the ATHENA vocabulary in which the resulting CDM concepts were then used to identify the childbirth delivery records in the EHR. Second, since these codes may not comprehensively capture all the OMOP CDM concepts indicating childbirth delivery, we explored the semantic relationships of the OMOP CDM concepts retrieved by these codes and supplemented them with the newly identified concepts.[26] The final concept set contained 105 OMOP CDM standard concepts (See Appendix 3). Researchers (TL and YS) performed the phenotyping and were not involved in the phenotyping evaluation.

*Rule-based algorithm.* Upon manual chart review of the EHR data, we found the OMOP CDM concepts with a domain type of procedure have the highest accuracy with respect to determining the DOD, followed by domain types of condition and then observation. Thus, we developed a rule-based algorithm to approximate the true DOD by prioritizing the OMOP CDM 'procedure' domain over the 'condition' domain over the 'observation' domain (Figure 3). Table 3 shows the pseudocode for the algorithm.

All data manipulation, phenotyping, and algorithms were implemented using SQL, R, and PySpark on the N3C platform. Source programming codes are available at N3C, project "[RP-2B9622] Assessing and predicting the clinical outcomes of pregnant women with COVID-19 using machine learning approach."

### *Evaluation*

We performed a multi-level evaluation to test the validity of the algorithm as well as inter-rater reliability. To test the content validity of the OMOP CDM concepts resulting from the phenotyping, two researchers (CL and NG) independently reviewed the concept IDs and their semantic meanings and properties on the ATHENA vocabulary and rated dichotomously on the relevance of all concept IDs. Inter-rater reliability was measured by Cohen's Kappa. Disagreements were discussed and resolved together with a senior OB/GYN physician (BC).

To test the clinical validity of the algorithm for inferring GA, we randomly selected 30 patients from the final cohort, which resulted in 40 distinct gestations, including multiple gestations. Their comprehensive medical records on GA, excluding laboratory data, were pulled from the EHR. We calculated the start date of pregnancy by subtracting the GA in weeks from the event date for each record. Two clinical experts (CL and NG) independently reviewed the retrieved records and rated them based on two metrics: accuracy (high/moderate/low) and granularity (high/moderate/low).

Accuracy is concerned with the level that the selected OMOP CDM concepts can accurately indicate the GA. For example, GA-related records are typically documented during antenatal care visits. Multiple GA-related records, even when documented at different dates, can suggest a consistent GA. The algorithm-selected concept is most accurate if it is among these records. When there are other GA-related records suggesting a GA different from the algorithm-selected record, the accuracy would not be high. Granularity refers to the extent that the algorithm-selected concept can indicate a specific gestational week. For example, the "gestation period, 38 weeks" has a high granularity level whereas "third-trimester pregnancy" has a low granularity level.

To test the clinical validity of the algorithm for inferring DOD, we randomly selected 30 gestations from the final cohort. Their records consisting of procedures, conditions, observations, measurements were pulled from the EHR within ± 14 days of estimated DOD. Two clinical experts (CL and NG) independently reviewed the charts and labeled whether the DOD was correctly inferred by the algorithm.

Despite the average gestation being around 280 days, this estimation varies among individuals. To represent rare cases such as preterm birth, post-term birth, and early-stage termination, we also performed extreme value analysis, in which two clinical experts (TL and CL) performed chart review for 30 randomly selected samples with <150 or >300 days of gestation.

### *Characteristics of pregnant women with and without COVID-19*

Using TED-PC, we performed descriptive analyses to explore maternal demographics and underlying conditions for pregnant women with (cases) and without COVID-19 (controls) which are characterized by temporal information of the gestational weeks when SARS-CoV-2 infection was identified.

## Results

### *Identified OMOP CDM concepts*

We identified 2,773 OMOP CDM concepts from the ATHENA vocabulary, of which 2,370 indicated the patient being pregnant. Among the concepts relating to pregnancy, 336 have GA-related information. We excluded 189 concepts that either indicated the inaccurate time range broader than one trimester (13 weeks) (e.g., concept ID 21493940: US for pregnancy in the second or third trimester) or did not have corresponding records in the N3C database (e.g., concept ID 3025286: Gestational age estimated from foot length on US by Mercer 1987 method). Totally, 138 concepts contained useful gestational week information with a time range from one week to one trimester. Within the selected 138 concepts, 42 were in high accuracy, 9 were in moderate-high accuracy, 5 were in moderate-low accuracy, and 82 were in low accuracy.

### *Algorithm performance*

To evaluate phenotyping results, the content validity of the selected concepts was assessed and rated blindfolded by two independent reviewers who did not participate in the phenotyping. Both reviewers rated all concepts as "valid" (100% agreement).

We evaluated the performance of the GA algorithm in two dimensions: accuracy and granularity. Among the 30 randomly selected mothers, eight of them had two gestations and one of them had three gestations during the study time frame. The mean gestation length was 270.15 days with a maximum of 299 days and a minimum of 159 days. Among the 40 pregnancies, one reviewer rated 34 (85.0%) samples as high accuracy, 4 (10.0%) samples as moderate accuracy, and 2 (5.0%) sample as low accuracy. The other reviewer rated 35 (87.5%) samples as high accuracy, 3 (7.5%) samples as moderate accuracy, and 2 (5.0%) samples as low accuracy. The Cohen's Kappa with linear weighting is 0.62, CI = [0.35, 0.90]. For granularity, both reviewers rated the 39 samples as high granularity and one sample as low granularity (100% agreement, unweighted Cohen's Kappa = 1). See Table 4 for the confusion matrix.

For the DOD algorithm, a total of 30 patients' EHR were reviewed independently. Both reviewers rated the 30 samples to be accurate (100% agreement, unweighted Cohen's Kappa = 1)

***Extreme value analysis***

We randomly selected 30 gestations with the gestation length either smaller than 150 days or greater than 300 days and extracted their EHR. After chart review, 28 of the 30 gestations were extracted with correct GA information, with an accuracy of 93.3%. For the two error cases, the first one was due to the contradiction between the GA records on different dates. For example, there was a record with the concept name "Gestation period, 38 weeks" on date 1, but other records with the same concept on date 2. The second error case was due to the contradiction between the GA records on the same date. For example, on the same date, one record was with concept name "Gestation period, 36 weeks" and another record was with concept name "Gestation period, 39 weeks". Among the correctly inferred cases, a certain level of inaccuracy existed. Five gestations only had low accuracy level GA records. Among the extreme case, one of them had only one GA record. Inter-rater reliability is 100% (unweighted Cohen's Kappa = 1).

***Characteristics of pregnant women with and without COVID-19***

Between 2018-06-01 and 2021-05-31, a total of 296,194 gestations in 270,897 pregnant women were identified from the N3C database. The mean and the median ages are 30.31 and 31, respectively. There were 245,892 women who had one pregnancy during the study time, 24,713 and 292 had two and three pregnancies, respectively. The mean gestation length was 274.14 days. The median was 278 days, with a minimum of 140 days and a maximum of 308 days. N3C data retrieval was completed on 02/12/2022.

Using TED-PC, we identified the timing of SARS-CoV-2 infections in gestational weeks. Figure 4 shows the frequency of infections across gestational weeks. More than half of the infections happened during late pregnancy (between 32 and 41 week), which might be related to the increased antenatal visits in late pregnancy. Table 5 provides the selected demographics and underlying conditions of the cohort captured by TED-PC, stratified by trimesters. There were 104,791 and 191,403 gestations before and during the COVID-19 pandemic, respectively, among which there are 16,659 gestations with COVID-19 and 174,744 without COVID-19 peri-pandemic. Age group 30-34 shared the largest proportion across the age groups, followed by the age groups 25-29, 40-44, and 20-24. White people made up the largest percentage of nearly 50% of the total population, followed by Black and Hispanic/Latino races. For mothers who had ever been infected by SARS-CoV-2 before the DOD (before or during the pregnancy), age groups 30-34, 20-24, and 25-29 had the largest percentage. Besides, the percentage of the White was smaller (39.7%) compared with that of pre-pandemic (57.1%). Hispanic/Latino made up the largest proportion across the races in those SARS-CoV-2 infected mothers (31.5%). Compared with the pregnant women without COVID-19, pregnant women with COVID-19 had a higher prevalence in obesity (35.1% vs. 29.5%), diabetes (17.8% vs. 17.0%), chronic obstructive pulmonary disease (COPD) (0.2% vs. 0.1%), respiratory distress syndrome (ARDS) (1.8% vs. 0.2%), myocardial infarction (0.2% vs.

0.1%), and HIV/AIDS (0.6% vs. 0.4%). The characteristics of the proportions shared similar trends when stratified by different trimesters.

## Discussion

Using N3C data, we created the first EHR-based cohort of SARS-CoV-2-infected pregnant women with complete temporal information of clinical events spanning the gestation length, which supports urgently needed COVID-19 research for pregnant women in the US. Our algorithm is among the first that can detect temporal information of pregnancy care including early-state pregnancy, preterm birth, early termination, and post-term birth. This algorithm shows the promise to underpin EHR deep phenotyping of pregnancy care as well as machine learning methods that require precise temporal information of clinical events. As a rapid development of clinical information extraction tool for combating COVID-19, our algorithm is currently supporting several EHR-based cohort studies on the N3C to examine the impact of COVID-19 on pregnant women's real-time clinical inflammatory progression and pregnancy complications.

The accuracy of the TED-PC is warranted by a few logic layers. First, compared with the previous studies that focused on the claims data or required labor efforts, our study took advantage of the OMOP CDM to normalize EHR data and categorized normalized concepts into different priority groups.[15,20-22,27-30] For example, the GA algorithm prioritized the concepts in the "Procedure" domain over "Condition" domain over "Observation" domain, which logically prevented the algorithm from selecting the records that were at a higher risk of semantic ambiguity and low granularity. Second, our algorithm categorized the GA-related concepts into different accuracy levels by indicating time range. This step allows the TED-PC to prioritize the records with the most accurate information. Third, the 270-day interval in our algorithm enabled us to distinguish different gestations of the same mother within the time frame for both the estimations of GA and DOD. Fourth, the merging and matching process of the GA cohort and the DOD cohort can exclude the gestations with untrustworthy or missing values that are due to incomplete EHR data.

Detection of temporal information for pregnant women with COVID-19 is made available by using N3C data for its two major features. First, N3C has multi-system EHR data linked at individual level. This unique feature enables our algorithm to impute a huge amount of missing temporal values and to resolve conflicts of temporal values among health records from different hospital systems. Second, N3C includes annotated clinical notes, procedures, and laboratory tests/results that are normalized with OMOP CDM, which allow the algorithms to leverage multi-source contextual information for inferring temporal information at an adequate level of granularity.

A few limitations of this study warrant to note. First, because several GA-related OMOP CDM concepts do not indicate specific gestational weeks (e.g., "Spontaneous onset of labor between 37 and 39 week gestation with planned cesarean section"), we inferred the gestational weeks using the median time point of the range, which may impair the performance of the algorithm. This impact is mild on the concepts with high or moderate-high accuracy levels since the time range is small, but it could be severe in the concepts with the low accuracy level. Second, our EHR data may not be comprehensive. For example, some examinations or laboratory tests do not have time information, but they are often prescribed to the mothers during a specific time frame of gestation. Our future direction will aim to improve the performance of TED-PC and test the external validity. From error analysis, data incompleteness and inconsistency remain the major sources of error. Well-designed EHR data imputation methods and a hybrid model of rule-based and machine learning algorithms hold promises for addressing these issues. Although our algorithm is designed for N3C data, it could be potentially repurposed for other OMOP CDM normalized EHR.

## Conclusion

We explored and compared the characteristics of pregnant women by different timing of SARS-CoV-2 infection with our newly developed technique: TED-PC, a rule-based algorithm to automatically infer comprehensive temporal information of clinical events from EHR during the pregnancy care. The performance of TED-PC is satisfactory as collectively, accuracy and granularity of temporal information are beyond 90%. TED-PC has been implemented on N3C, supporting multiple national EHR cohorts for desperately needed research on the impact of COVID-19 on pregnancy. TED-PC is designed for N3C data but remains generalizable for OMOP CDM normalized EHR.

## Acknowledgment

This study is sponsored by NIH/NIAID under award 3R01AI127203-05S2. We thank Ms. Ashlee Kim for her support of medical coding.

### *N3C Attribution*

The analyses described in this publication were conducted with data or tools accessed through the NCATS N3C Data Enclave https://covid.cd2h.org and N3C Attribution & Publication Policy v 1.2-2020-08-25b supported by NCATS U24 TR002306. This research was possible because of the patients whose information is included within the data and the organizations (https://ncats.nih.gov/n3c/resources/data-contribution/data-transfer-agreement-signatories) and scientists who have contributed to the on-going development of this community resource (https://doi.org/10.1093/jamia/ocaa196).

### *IRB*

The N3C data transfer to NCATS is performed under a Johns Hopkins University Reliance Protocol # IRB00249128 or individual site agreements with NIH. The N3C Data Enclave is managed under the authority of the NIH; information can be found at https://ncats.nih.gov/n3c/resources.

### *Individual Acknowledgements for Core Contributors*

We gratefully acknowledge contributions from the following N3C core teams (leads designated with asterisks):

• CD2H Principal Investigators and N3C Lead Investigators: Melissa A. Haendel*, Christopher G. Chute*, Anita Walden

• NCATS CD2H and N3C Science Officer: Kenneth R. Gersing

• NCATS CD2H and N3C Program Officer: Leonie Misquitta

• NCATS N3C Leadership Team: Joni L. Rutter*, Kenneth R. Gersing*, Penny Wung Burgoon, Samuel Bozzette, Mariam Deacy, Christopher Dillon, Rebecca Erwin-Cohen, Nicole Garbarini, Valery Gordon, Michael G. Kurilla, Emily Carlson Marti, Sam G. Michael, Leonie Misquitta, Lili Portilla, Clare Schmitt, Meredith Temple-O'Connor

• Workstream, subgroup and administrative leaders: Melissa A. Haendel*, Tellen D. Bennett, Christopher G. Chute, David A. Eichmann, Justin Guinney, Warren A. Kibbe, Hongfang Liu, Philip R.O. Payne, Emily R. Pfaff, Peter N. Robinson, Joel H. Saltz, Heidi Spratt, Justin Starren, Christine Suver, Adam B. Wilcox, Andrew E. Williams, Chunlei Wu

• Key liaisons at data partner sites

• Regulatory staff at data partner sites

• Individuals at the sites who are responsible for creating the datasets and submitting data to N3C

• Data Ingest and Harmonization Team: Christopher G. Chute*, Emily R. Pfaff*, Davera Gabriel, Stephanie S. Hong, Kristin Kostka, Harold P. Lehmann, Richard A. Moffitt, Michele Morris, Matvey B. Palchuk, Xiaohan Tanner Zhang, Richard L. Zhu

• Phenotype Team (Individuals who create the scripts that the sites use to submit their data, based on the COVID and Long COVID definitions): Emily R. Pfaff*, Benjamin Amor, Mark M. Bissell, Marshall Clark, Andrew T. Girvin, Stephanie S. Hong, Kristin Kostka, Adam M. Lee, Robert T. Miller, Michele Morris, Matvey B. Palchuk, Kellie M. Walters

• N3C Community Project Management and Operations Team: Anita Walden*, Will Cooper, Patricia A. Francis, Rafael Fuentes, Alexis Graves, Julie A. McMurry, Andrew J. Neumann, Shawn T. O'Neil, Usman Sheikh, Elizabeth Zampino

• Analytics Team (Individuals who build the Enclave infrastructure, help create codesets, variables, and help Domain Teams and project teams with their datasets): Benjamin Amor*, Mark M. Bissell, Katie Rebecca Bradwell, Andrew T. Girvin, Amin Manna, Nabeel Qureshi

• Publication Committee Team: Mary Morrison Saltz*, Christine Suver*, Christopher G. Chute, Melissa A. Haendel, Julie A. McMurry, Andréa M. Volz, Anita Walden, Carolyn Bramante, Jeremy Richard Harper, Wenndy Hernandez, Farrukh M Koraishy, Federico Mariona, Amit Saha, Satyanarayana Vedula

Individual researchers who developed OMOP concept sets that have been used in this study: Daniel Meza, Arti Patel, Richard Zhu, Alfred Anzalone, Arti Patel, and Benjamin Amor.

We acknowledge support from many grants; the content is solely the responsibility of the authors and does not necessarily represent the official views of the N3C Program, the NIH or other funders. In addition, access to N3C Data Enclave resources does not imply endorsement of the research project and/or results by NIH or NCATS.

***Data Partners with Released Data***

The following institutions whose data is released or pending:

Available: Advocate Health Care Network — UL1TR002389: The Institute for Translational Medicine (ITM) • Boston University Medical Campus — UL1TR001430: Boston University Clinical and Translational Science Institute • Brown University — U54GM115677: Advance Clinical Translational Research (Advance-CTR) • Carilion Clinic — UL1TR003015: iTHRIV Integrated Translational health Research Institute of Virginia • Charleston Area Medical Center — U54GM104942: West Virginia Clinical and Translational Science Institute (WVCTSI) • Children's Hospital Colorado — UL1TR002535: Colorado Clinical and Translational Sciences Institute • Columbia University Irving Medical Center — UL1TR001873: Irving Institute for Clinical and Translational Research • Duke University — UL1TR002553: Duke Clinical and Translational Science Institute • George Washington Children's Research Institute — UL1TR001876: Clinical and Translational Science Institute at Children's National (CTSA-CN) • George Washington University — UL1TR001876: Clinical and Translational Science Institute at Children's National (CTSA-CN) • Indiana University School of Medicine — UL1TR002529: Indiana Clinical and Translational Science Institute • Johns Hopkins University — UL1TR003098: Johns Hopkins Institute for Clinical and Translational Research • Loyola Medicine — Loyola University Medical Center • Loyola University Medical Center — UL1TR002389: The Institute for Translational Medicine (ITM) • Maine Medical Center — U54GM115516: Northern New England Clinical & Translational Research (NNE-CTR) Network • Massachusetts General Brigham — UL1TR002541: Harvard Catalyst • Mayo Clinic Rochester — UL1TR002377: Mayo Clinic Center for Clinical and Translational Science (CCaTS) • Medical University of South Carolina — UL1TR001450: South Carolina Clinical & Translational Research Institute (SCTR) • Montefiore Medical Center — UL1TR002556: Institute for Clinical and Translational Research at Einstein and Montefiore • Nemours — U54GM104941: Delaware CTR ACCEL Program • NorthShore University HealthSystem — UL1TR002389: The Institute for Translational Medicine (ITM) • Northwestern University at Chicago — UL1TR001422: Northwestern University Clinical and Translational Science Institute (NUCATS) • OCHIN — INV-018455: Bill and Melinda Gates Foundation grant to Sage Bionetworks • Oregon Health & Science University — UL1TR002369: Oregon Clinical and Translational Research Institute • Penn State Health Milton S. Hershey Medical Center — UL1TR002014: Penn State Clinical and Translational Science Institute • Rush University Medical Center — UL1TR002389: The Institute for Translational Medicine (ITM) • Rutgers, The State University of New Jersey — UL1TR003017: New Jersey Alliance for Clinical and Translational Science • Stony Brook University — U24TR002306 • The Ohio State University — UL1TR002733: Center for Clinical and Translational Science • The State University of New York at Buffalo — UL1TR001412: Clinical and Translational Science Institute • The University of Chicago — UL1TR002389: The Institute for Translational Medicine (ITM) • The University of Iowa — UL1TR002537: Institute for Clinical and Translational Science • The University of Miami Leonard M. Miller School of Medicine — UL1TR002736: University of Miami Clinical and Translational Science Institute • The University of Michigan at Ann Arbor — UL1TR002240: Michigan Institute for Clinical and Health Research • The University of Texas Health Science Center at Houston — UL1TR003167: Center for Clinical and Translational Sciences (CCTS) • The University of Texas Medical Branch at Galveston — UL1TR001439: The Institute for Translational Sciences • The University of Utah — UL1TR002538: Uhealth Center for Clinical and Translational Science • Tufts Medical Center — UL1TR002544: Tufts Clinical and Translational Science Institute • Tulane University — UL1TR003096: Center for Clinical and Translational Science • University Medical Center New Orleans — U54GM104940: Louisiana Clinical and Translational Science (LA CaTS) Center • University of Alabama at Birmingham — UL1TR003096: Center for Clinical and Translational Science • University of Arkansas for Medical Sciences — UL1TR003107: UAMS Translational Research Institute • University of Cincinnati — UL1TR001425: Center for Clinical and Translational Science and Training • University of Colorado Denver, Anschutz Medical Campus — UL1TR002535: Colorado Clinical and Translational Sciences Institute • University of Illinois at Chicago — UL1TR002003: UIC Center for Clinical and Translational Science • University of Kansas Medical Center — UL1TR002366: Frontiers: University of Kansas Clinical and Translational Science Institute • University of Kentucky — UL1TR001998: UK Center for Clinical and Translational Science • University of Massachusetts Medical School Worcester — UL1TR001453: The UMass Center for Clinical and Translational Science (UMCCTS) • University of Minnesota — UL1TR002494: Clinical and Translational Science Institute • University of Mississippi Medical Center — U54GM115428: Mississippi Center for Clinical and Translational Research (CCTR) • University of Nebraska Medical Center — U54GM115458: Great Plains IDeA-Clinical & Translational Research • University of North Carolina at Chapel Hill — UL1TR002489: North Carolina Translational and Clinical Science Institute • University of Oklahoma Health Sciences Center — U54GM104938: Oklahoma Clinical and Translational Science Institute (OCTSI) • University of Rochester — UL1TR002001: UR Clinical & Translational Science Institute • University of Southern California — UL1TR001855: The Southern California Clinical and Translational Science Institute (SC CTSI) • University of Vermont — U54GM115516: Northern New England Clinical & Translational Research (NNE-CTR) Network • University of Virginia — UL1TR003015: iTHRIV Integrated Translational health Research Institute of Virginia • University of Washington — UL1TR002319: Institute of Translational Health Sciences • University of Wisconsin-Madison — UL1TR002373: UW Institute for Clinical and Translational Research • Vanderbilt University Medical Center — UL1TR002243: Vanderbilt Institute for Clinical and Translational Research • Virginia Commonwealth University — UL1TR002649: C. Kenneth and Dianne Wright Center for Clinical and Translational Research • Wake Forest University Health Sciences — UL1TR001420: Wake Forest Clinical and Translational Science Institute • Washington University in St. Louis — UL1TR002345: Institute of Clinical and Translational Sciences • Weill Medical College of Cornell University — UL1TR002384: Weill Cornell Medicine Clinical and Translational Science Center • West Virginia University — U54GM104942: West Virginia Clinical and Translational Science Institute (WVCTSI)

Submitted: Icahn School of Medicine at Mount Sinai — UL1TR001433: ConduITS Institute for Translational Sciences • The University of Texas Health Science Center at Tyler — UL1TR003167: Center for Clinical and Translational Sciences (CCTS) • University of California, Davis — UL1TR001860: UCDavis Health Clinical and Translational Science Center • University of California, Irvine — UL1TR001414: The UC Irvine Institute for Clinical and Translational Science (ICTS) • University of California, Los Angeles — UL1TR001881: UCLA Clinical Translational Science Institute • University of California, San Diego — UL1TR001442: Altman Clinical and Translational Research Institute • University of California, San Francisco — UL1TR001872: UCSF Clinical and Translational Science Institute

Pending: Arkansas Children's Hospital — UL1TR003107: UAMS Translational Research Institute • Baylor College of Medicine — None (Voluntary) • Children's Hospital of Philadelphia — UL1TR001878: Institute for Translational Medicine and Therapeutics • Cincinnati Children's Hospital Medical Center — UL1TR001425: Center for Clinical and Translational Science and Training • Emory University — UL1TR002378: Georgia Clinical and Translational Science Alliance • HonorHealth — None (Voluntary) • Loyola University Chicago — UL1TR002389: The Institute for Translational Medicine (ITM) • Medical College of Wisconsin — UL1TR001436: Clinical and Translational Science Institute of Southeast Wisconsin • MedStar Health Research Institute — UL1TR001409: The Georgetown-Howard Universities Center for Clinical and Translational Science (GHUCCTS) • MetroHealth — None (Voluntary) • Montana State University — U54GM115371: American Indian/Alaska Native CTR • NYU Langone Medical Center — UL1TR001445: Langone Health's Clinical and Translational Science Institute • Ochsner Medical Center — U54GM104940: Louisiana Clinical and Translational Science (LA CaTS) Center • Regenstrief Institute — UL1TR002529: Indiana Clinical and Translational Science Institute • Sanford Research — None (Voluntary) • Stanford University — UL1TR003142:

**International Committee of Medical Journal Editors (ICMJE) Statement:**

Authorship was determined using ICMJE recommendations.

**Authorship Contribution Statement**

Conceived study design: TL, CL

Developed methodology: TL, CL

Performed data analysis and interpretation: TL, CL, YS, BC, NG

Drafted manuscript: TL, CL

Reviewed, edited, and approved manuscript: TL, CL, JL, BC, PH, YS, NG, XL

**Declaration of Conflict of Interest**

None.

Table 1. Concepts Categorization by Accuracy Level

| Accuracy level | Time Interval | Definition | Example Concept IDs |
|---|---|---|---|
| High | 1 week | The concept name specifies the value of GA in week (e.g., Gestation period, 15 weeks). | 4337360: Gestation period, 1 week<br>4097608: Gestation period, 18 weeks<br>444098: Gestation period, 40 weeks |
| Moderate - High | 2-5 weeks | The concept name does not specify the value of GA in week but specifies the range of GA in week which is larger than 1 week and smaller than 6 weeks (e.g., Gestation 9-13 weeks) | 4181468: Gestation 9- 13 weeks<br>44791171: 9-13 weeks gestational age<br>45757118: Spontaneous onset of labor between 37 and 39 weeks gestation with planned cesarean section |
| Moderate - Low | 6-10 weeks | The concept name does not specify the value of GA in week but specifies the range of GA in week which is larger than 5 week and smaller than 11 weeks (e.g., Gestation 14-20 weeks) | 4180111: Third trimester pregnancy less than 36 weeks<br>4178165: Gestation 14-20 weeks<br>44791170: 14-20 weeks gestational age |
| Low | 11-13 weeks | The concept name does not specify the value of GA in week but specifies the trimester (e.g., first trimester) | 3657563: First trimester bleeding<br>4239938: First trimester pregnancy<br>4112238: Third trimester |

Table 2. The Pseudocode for the Algorithm: Estimating the Gestational Age

**Algorithm:** GA estimation for each gestation

```
procedure GA ESTIMATION
    Input: GA-related clinical events, the date of GA-related clinical events
    Output: Estimated GA for each gestation
    pregnancy date <- calculated by the date of GA-related clinical events
    if the time range indicated by the GA-related clinical events == 1 week then
        accuracy <- 1
    else if 2 weeks <= the time range indicated by the GA-related clinical events <= 5 weeks then
        accuracy <- 2
    else if 6 weeks <= the time range indicated by the GA-related clinical events <= 10 weeks then
        accuracy <- 3
    else if 11 weeks <= the time range indicated by the GA-related clinical events <= 13 weeks then
        accuracy <- 4
    endif
    for each patient
        Sort the GA-related clinical events by event date chronologically
        repeat
            Select the first pregnancy date with the highest accuracy*
            for the GA-related clinical events in (±270 days of the selected pregnancy date)
                Remove
            endfor
            Estimated GA <- the first GA-related clinical event with the highest accuracy*
        until the last GA-related clinical event
        repeat
            if remaining GA-related clinical events exist then
                repeat
                    Select the first pregnancy date with the highest accuracy*
                    for the GA-related clinical events in (±270 days of the selected pregnancy date)
                        Remove
                    endfor
                    Estimated GA <- the first GA-related clinical event with the highest accuracy*
                until the GA-related clinical event
            else stop
            endif
        until the GA-related clinical event
    endfor
endprocedure
```

* Accuracy level: 1-4, from high to low

Table 3. The Pseudocode for the Algorithm: Estimating the Date of Delivery

**Algorithm:** DOD estimation for each gestation

```
procedure DOD ESTIMATION
    Input: childbirth delivery clinical events, the domain and the date of DOD-related clinical events
    Output: Estimated DOD for each gestation
    Event DOD <- the date of childbirth delivery clinical events
    if domain == "procedure" then
        accuracy <- 1
    else if domain == "condition" then
        accuracy <- 2
    else if domain == "observation" then
        accuracy <- 3
    endif
    for each patient
        Sort the childbirth delivery clinical events by event DOD in reversed chronological order
        repeat
            Select the first event DOD with the highest accuracy*
            for the childbirth delivery clinical event in (±270 days of the selected event DOD)
                Remove
            endfor
            Estimated DOD <- the first event DOD with the highest accuracy*
        until the last childbirth delivery clinical event
        repeat
            if remaining childbirth delivery clinical events exist then
                repeat
                    Select the first event DOD with the highest accuracy*
                    for the childbirth delivery clinical events in (±270 days of the selected event DOD)
                        Remove
                    endfor
                    Estimated DOD <- the first event DOD with the highest accuracy*
                until the last childbirth delivery clinical event
            else stop
            endif
        until the last childbirth delivery clinical event
    endfor
endprocedure
```

* Accuracy level: 1-3, from high to low

Table 4. The Confusion Matrix of the Accuracy Rating for the Performance of the GA Algorithm

| | | Reviewer2 | | | |
|---|---|---|---|---|---|
| | | High | Moderate | Low | Total |
| Reviewer1 | High | 33 | 1 | 0 | 34 |
| | Moderate | 2 | 1 | 1 | 4 |
| | Low | 0 | 1 | 1 | 2 |
| | Total | 35 | 3 | 2 | 40 |

Table 5. Selected Demographics and Underlying Conditions for Pregnant Women With and Without COVID-19 by Gestations During Pre- and Peri-Pandemic

| | **Pre-Pandemic**[*] | | **Peri-Pandemic**[†] | | | | | | | | | | | | | |
|---|---|---|---|---|---|---|---|---|---|---|---|---|---|---|---|---|
| | **Total** | | **Total** | | **COVID-19 Before the DOD** | | | | **COVID-19 During the First or Second Trimester (before 27 weeks)** | | | | **COVID-19 During the Third Trimester (after 28 weeks)**[‡] | | | |
| **Characteristics** | n=104,791 | | n=191,403 | | No (n=174,744) | | Yes (n=16,659) | | No (n=187,351) | | Yes (n=4,052) | | No (n=178,449) | | Yes (n=9,327) | |
| **Age group** | | | | | | | | | | | | | | | | |
| **15-19** | 3,296 | 3.1% | 5,366 | 2.8% | 4,778 | 2.7% | 588 | 3.5% | 5,268 | 2.8% | 98 | 2.4% | 4,918 | 2.8% | 354 | 3.8% |
| **20-24** | 17,076 | 16.3% | 28,000 | 14.6% | 25,051 | 14.3% | 2,949 | 17.7% | 27,345 | 14.6% | 655 | 16.2% | 25,796 | 14.5% | 1,663 | 17.8% |
| **25-29** | 27,337 | 26.1% | 46,867 | 24.5% | 42,406 | 24.3% | 4,461 | 26.8% | 45,783 | 24.4% | 1,084 | 26.8% | 43,487 | 24.4% | 2,522 | 27.0% |
| **30-34** | 33,590 | 32.1% | 60,815 | 31.8% | 56,057 | 32.1% | 4,758 | 28.6% | 59,606 | 31.8% | 1,209 | 29.8% | 57,105 | 32.0% | 2,671 | 28.6% |
| **35-39** | 18,843 | 18.0% | 39,162 | 20.5% | 36,173 | 20.7% | 2,989 | 17.9% | 38,395 | 20.5% | 767 | 18.9% | 36,713 | 20.6% | 1,621 | 17.4% |
| **40-44** | 4,326 | 4.1% | 10,402 | 5.4% | 9,553 | 5.5% | 849 | 5.1% | 10,179 | 5.4% | 223 | 5.5% | 9,698 | 5.4% | 458 | 4.9% |
| **45-49** | 323 | 0.3% | 791 | 0.4% | 726 | 0.4% | 65 | 0.4% | 775 | 0.4% | - | - | 732 | 0.4% | 38 | 0.4% |
| **Race** | | | | | | | | | | | | | | | | |
| **White** | 59,842 | 57.1% | 95,517 | 49.9% | 88,898 | 50.9% | 6,619 | 39.7% | 93,736 | 50.0% | 1,781 | 44.0% | 90,397 | 50.7% | 3,660 | 39.2% |
| **Black** | 18,536 | 17.7% | 33,144 | 17.3% | 30,567 | 17.5% | 2,577 | 15.5% | 32,511 | 17.4% | 633 | 15.6% | 30,740 | 17.2% | 1,412 | 15.1% |
| **Hispanic/Latino** | 15,961 | 15.2% | 37,095 | 19.4% | 31,841 | 18.2% | 5,254 | 31.5% | 35,964 | 19.2% | 1,131 | 27.9% | 33,249 | 18.6% | 3,149 | 33.8% |
| **Asian** | 4,219 | 4.0% | 9,952 | 5.2% | 9,289 | 5.3% | 663 | 4.0% | 9,784 | 5.2% | 168 | 4.1% | 9,447 | 5.3% | 348 | 3.7% |
| **NHOPI** | 182 | 0.2% | 375 | 0.2% | 325 | 0.2% | 50 | 0.3% | 361 | 0.2% | - | - | 329 | 0.2% | 29 | 0.3% |
| **Other/unknown** | 5,477 | 5.2% | 13,996 | 7.3% | 12,643 | 7.2% | 1,353 | 8.1% | 13,703 | 7.3% | 293 | 7.2% | 13,051 | 7.3% | 673 | 7.2% |
| **Multiracial** | 574 | 0.5% | 1,324 | 0.7% | 1,181 | 0.7% | 143 | 0.9% | 1,292 | 0.7% | 32 | 0.8% | 1,236 | 0.7% | 56 | 0.6% |
| **Obesity** | | | | | | | | | | | | | | | | |
| **No** | 77,992 | 74.4% | 133,988 | 70.0% | 123,181 | 70.5% | 10,807 | 64.9% | 131,423 | 70.1% | 2,565 | 63.3% | 125,945 | 70.6% | 5,611 | 60.2% |
| **Yes** | 26,799 | 25.6% | 57,415 | 30.0% | 51,563 | 29.5% | 5,852 | 35.1% | 55,928 | 29.9% | 1,487 | 36.7% | 52,504 | 29.4% | 3,716 | 39.8% |
| **Hypertensive disorders (any)**[§] | | | | | | | | | | | | | | | | |
| **No** | 85,308 | 81.4% | 150,308 | 78.5% | 137,226 | 78.5% | 13,082 | 78.5% | 147,156 | 78.5% | 3,152 | 77.8% | 140,521 | 78.7% | 7,020 | 75.3% |
| **Yes** | 19,483 | 18.6% | 41,095 | 21.5% | 37,518 | 21.5% | 3,577 | 21.5% | 40,195 | 21.5% | 900 | 22.2% | 37,928 | 21.3% | 2,307 | 24.7% |
| **Diabetes (any)**[ǁ] | | | | | | | | | | | | | | | | |
| **No** | 90,059 | 85.9% | 158,651 | 82.9% | 144,962 | 83.0% | 13,689 | 82.2% | 155,379 | 82.9% | 3,272 | 80.8% | 148,036 | 83.0% | 7,509 | 80.5% |
| **Yes** | 14,732 | 14.1% | 32,752 | 17.1% | 29,782 | 17.0% | 2,970 | 17.8% | 31,972 | 17.1% | 780 | 19.2% | 30,413 | 17.0% | 1,818 | 19.5% |
| **COPD** | | | | | | | | | | | | | | | | |
| **No** | 104,683 | 99.9% | 191,120 | 99.9% | 174,499 | 99.9% | 16,621 | 99.8% | 187,077 | 99.9% | 4,043 | 99.8% | 178,200 | 99.9% | 9,301 | 99.7% |
| **Yes** | 108 | 0.1% | 283 | 0.1% | 245 | 0.1% | 38 | 0.2% | 274 | 0.1% | - | - | 249 | 0.1% | 26 | 0.3% |
| **ARDS** | | | | | | | | | | | | | | | | |
| **No** | 104,634 | 99.9% | 190,669 | 99.6% | 174,316 | 99.8% | 16,353 | 98.2% | 186,685 | 99.6% | 3,984 | 98.3% | 177,995 | 99.7% | 9,092 | 97.5% |
| **Yes** | 157 | 0.1% | 734 | 0.4% | 428 | 0.2% | 306 | 1.8% | 666 | 0.4% | 68 | 1.7% | 454 | 0.3% | 235 | 2.5% |
| **Myocardial Infarction** | | | | | | | | | | | | | | | | |
| **No** | 104,657 | 99.9% | 191,119 | 99.9% | 174,485 | 99.9% | 16,634 | 99.8% | 187,073 | 99.9% | 4,046 | 99.9% | 174,485 | 97.8% | 16,634 | 178.3% |
| **Yes** | 134 | 0.1% | 284 | 0.1% | 259 | 0.1% | 25 | 0.2% | 278 | 0.1% | - | - | 259 | 0.1% | 25 | 0.3% |
| **Congestive Heart Failure** | | | | | | | | | | | | | | | | |
| **No** | 104,395 | 99.6% | 190,634 | 99.6% | 174,045 | 99.6% | 16,589 | 99.6% | 186,600 | 99.6% | 4,034 | 99.6% | 178,195 | 99.9% | 9,309 | 99.8% |
| **Yes** | 396 | 0.4% | 769 | 0.4% | 699 | 0.4% | 70 | 0.4% | 751 | 0.4% | - | - | 254 | 0.1% | - | 0.2% |
| **HIV/AIDS** | | | | | | | | | | | | | | | | |
| **No** | 104,406 | 99.6% | 190,583 | 99.6% | 174,021 | 99.6% | 16,562 | 99.4% | 186,556 | 99.6% | 4,027 | 99.4% | 177,706 | 99.6% | 9,272 | 99.4% |
| **Yes** | 385 | 0.4% | 820 | 0.4% | 723 | 0.4% | 97 | 0.6% | 795 | 0.4% | 25 | 0.6% | 743 | 0.4% | 55 | 0.6% |
| **Placental Abruption** | | | | | | | | | | | | | | | | |
| **No** | 103,872 | 99.1% | 189,319 | 98.9% | 172,832 | 98.9% | 16,487 | 99.0% | 185,316 | 98.9% | 4,003 | 98.8% | 176,672 | 99.0% | 9,218 | 98.8% |
| **Yes** | 919 | 0.9% | 2,084 | 1.1% | 1,912 | 1.1% | 172 | 1.0% | 2,035 | 1.1% | 49 | 1.2% | 1,777 | 1.0% | 109 | 1.2% |

[*] Before the COVID-19 pandemic: June 01, 2018 to February 29, 2020

[†] During the COVID-19 pandemic: May 01, 2020 to May 31, 2021

[‡]: among those pregnancies with gestational length greater than 27 weeks

[§]: includes chronic hypertension and gestational hypertension

[ǀ]: includes chronic diabetes and gestational diabetes

DOD: date of delivery

NHOPI: Native Hawaiians and other Pacific Islanders

COPD: chronic obstructive pulmonary disease

ARDS: respiratory distress syndrome

-: Patient numbers < 20 were not reported

Figure 1. The Pipeline of TED-PC for Extracting Information and Building Cohorts

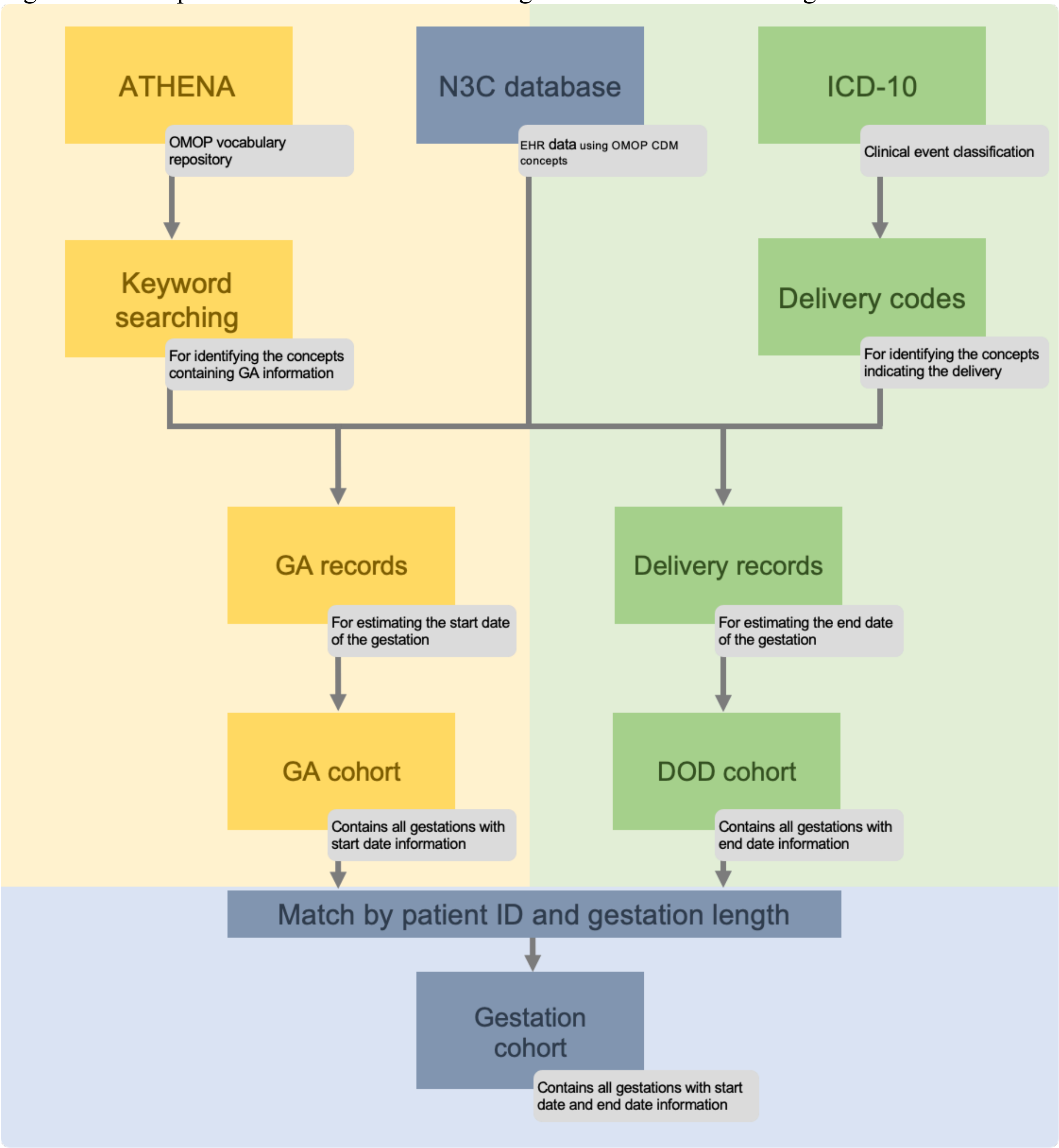

Figure 2. The Flowchart for Estimating Gestational Age and Pregnancy Date

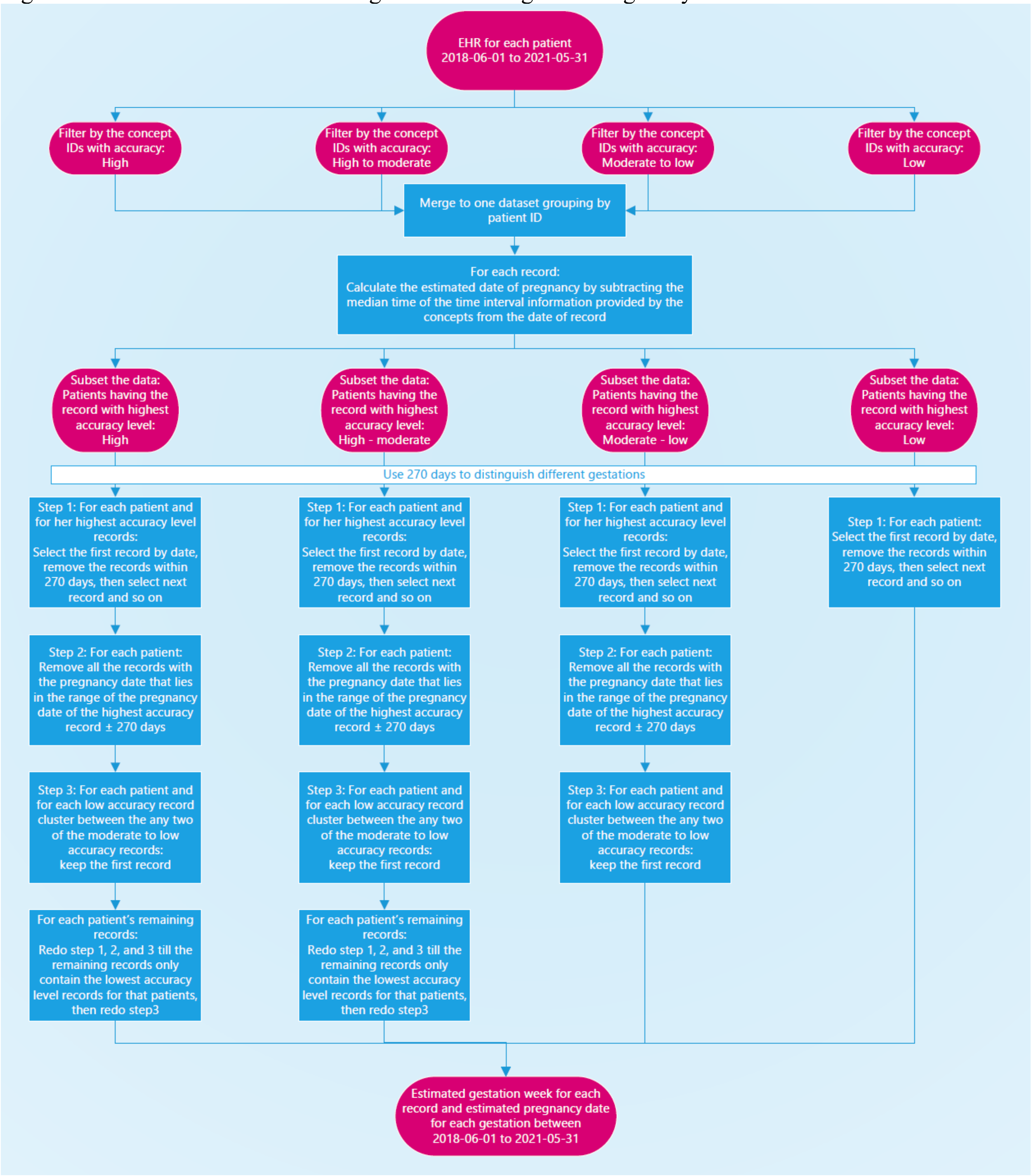

Figure 3. The Flowchart for Estimating Date of Delivery

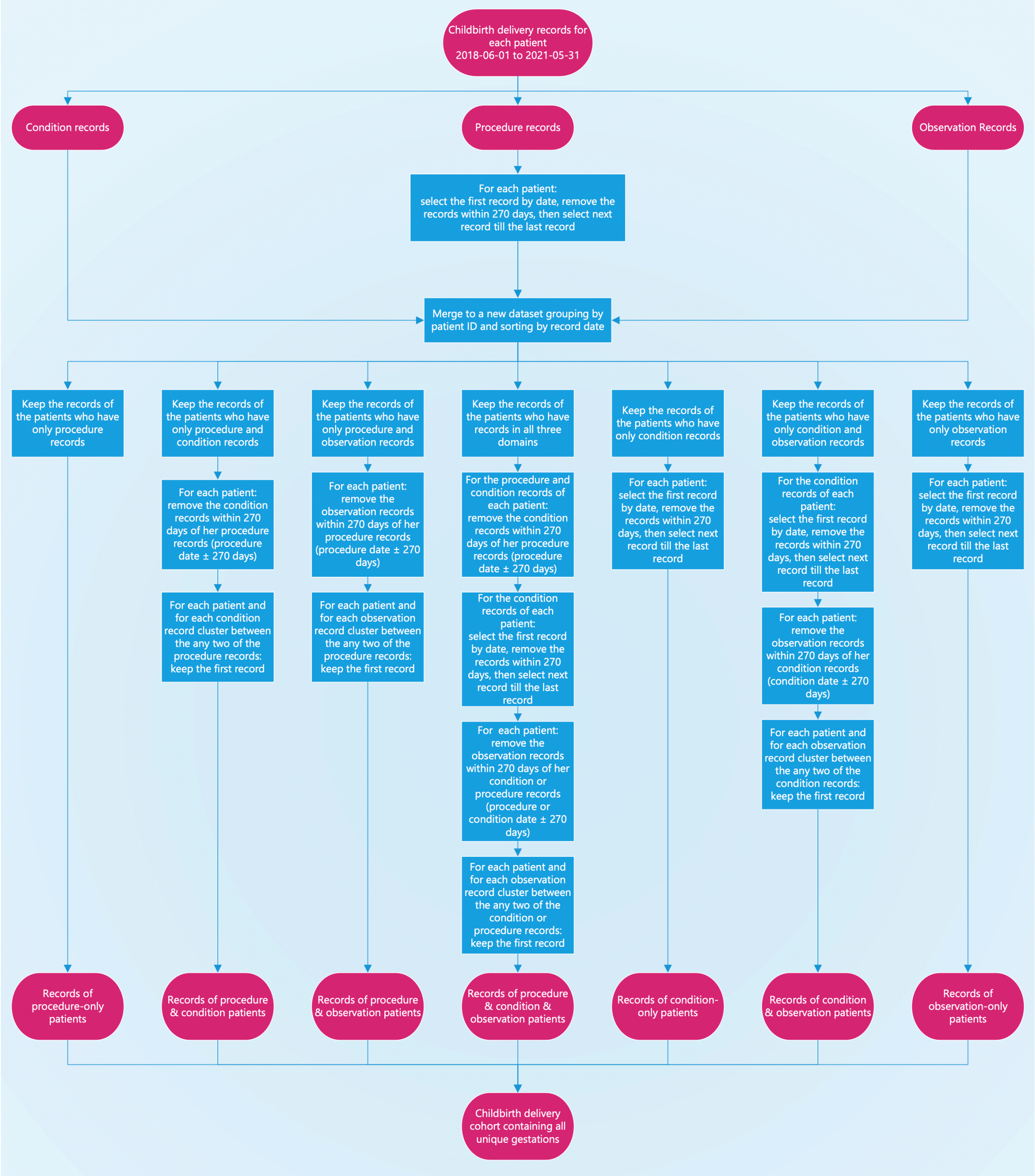

Figure 4. The Distribution of the SARS-CoV-2 Infection Cases by Gestational Week

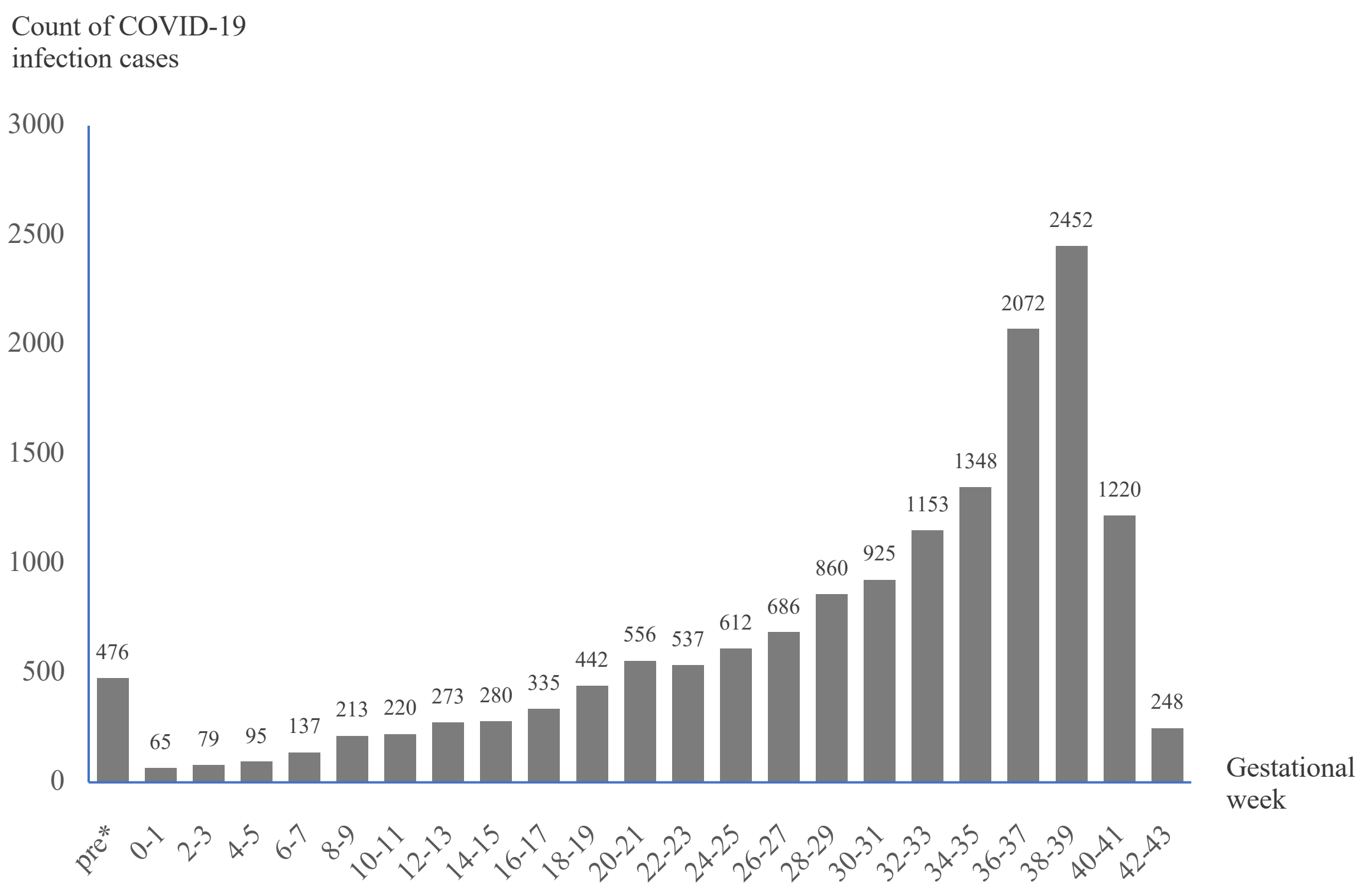

* pre-pregnancy

Appendix 1. OMOP CDM concepts for gestational age-related EHR records.

| Concept ID | Concept Name | Class | Domain | Vocabulary |
|---|---|---|---|---|
| 2793352 | Ultrasonography of Third Trimester, Multiple Gestation | ICD10PCS | Procedure | ICD10PCS |
| 2825091 | Imaging @ Fetus and Obstetrical @ Ultrasonography @ Third Trimester, Multiple Gestation | ICD10PCS Hierarchy | Procedure | ICD10PCS |
| 2871670 | Imaging @ Fetus and Obstetrical @ Ultrasonography @ Third Trimester, Multiple Gestation @ None | ICD10PCS Hierarchy | Procedure | ICD10PCS |
| 2898834 | Imaging @ Fetus and Obstetrical @ Ultrasonography @ Third Trimester, Multiple Gestation @ None @ None | ICD10PCS Hierarchy | Procedure | ICD10PCS |
| 4112238 | Third trimester | Qualifier Value | Observation | SNOMED |
| 4218813 | Third trimester pregnancy | Clinical Finding | Condition | SNOMED |
| 4142581 | Third trimester bleeding | Clinical Finding | Condition | SNOMED |
| 4167685 | Miscarriage in third trimester | Clinical Finding | Condition | SNOMED |
| 40487413 | Ultrasonography in third trimester | Procedure | Procedure | SNOMED |
| 4032055 | Threatened miscarriage in third trimester | Clinical Finding | Condition | SNOMED |
| 45757176 | Preterm labor in third trimester with preterm delivery in third trimester | Clinical Finding | Condition | SNOMED |
| 4029320 | Prenatal state of fetus, 3rd trimester | Clinical Finding | Condition | SNOMED |
| 35625971 | Vomiting during third trimester of pregnancy | Clinical Finding | Condition | SNOMED |
| 36713272 | Three dimensional obstetric ultrasonography in third trimester | Procedure | Procedure | SNOMED |
| 2793351 | Ultrasonography of Third Trimester, Single Fetus | ICD10PCS | Procedure | ICD10PCS |
| 3032525 | Fetal Narrative [Interpretation] Study observation.general 3rd trimester US | Clinical Observation | Measurement | LOINC |
| 3032291 | Fetal Narrative [Interpretation] Study observation.general 3rd trimester, multiple fetuses US | Clinical Observation | Measurement | LOINC |
| 2898832 | Imaging @ Fetus and Obstetrical @ Ultrasonography @ Third Trimester, Single Fetus | ICD10PCS Hierarchy | Procedure | ICD10PCS |
| 2898833 | Imaging @ Fetus and Obstetrical @ Ultrasonography @ Third Trimester, Single Fetus @ None | ICD10PCS Hierarchy | Procedure | ICD10PCS |
| 2807693 | Imaging @ Fetus and Obstetrical @ Ultrasonography @ Third Trimester, Single Fetus @ None @ None | ICD10PCS Hierarchy | Procedure | ICD10PCS |
| 2793350 | Ultrasonography of Second Trimester, Multiple Gestation | ICD10PCS | Procedure | ICD10PCS |
| 2832899 | Imaging @ Fetus and Obstetrical @ Ultrasonography @ Second Trimester, Multiple Gestation | ICD10PCS Hierarchy | Procedure | ICD10PCS |
| 2871669 | Imaging @ Fetus and Obstetrical @ Ultrasonography @ Second Trimester, Multiple Gestation @ None | ICD10PCS Hierarchy | Procedure | ICD10PCS |
| 2879607 | Imaging @ Fetus and Obstetrical @ Ultrasonography @ Second Trimester, Multiple Gestation @ None @ None | ICD10PCS Hierarchy | Procedure | ICD10PCS |
| 4113140 | Second trimester | Qualifier Value | Observation | SNOMED |
| 4327745 | Second trimester bleeding | Clinical Finding | Condition | SNOMED |
| 4244438 | Second trimester pregnancy | Clinical Finding | Condition | SNOMED |
| 40486918 | Ultrasonography in second trimester | Procedure | Procedure | SNOMED |
| 4224646 | Miscarriage in second trimester | Clinical Finding | Condition | SNOMED |
| 4294259 | Threatened miscarriage in second trimester | Clinical Finding | Condition | SNOMED |
| 45757175 | Preterm labor in second trimester with preterm delivery in second trimester | Clinical Finding | Condition | SNOMED |
| 4240362 | Prenatal state of fetus, 2nd trimester | Clinical Finding | Condition | SNOMED |
| 45763591 | Induced termination of pregnancy in second trimester | Clinical Finding | Condition | SNOMED |
| 40483600 | Measurement of alpha fetoprotein in second trimester | Procedure | Measurement | SNOMED |
| 4202200 | Surgical treatment of missed miscarriage of second trimester | Procedure | Procedure | SNOMED |
| 4290406 | Surgical treatment of missed miscarriage of third trimester | Procedure | Procedure | SNOMED |
| 2110327 | Treatment of missed abortion, completed surgically; second trimester | CPT4 | Procedure | CPT4 |
| 36305399 | Cigarettes smoked per day by Mother--2nd trimester | Survey | Observation | LOINC |
| 2793349 | Ultrasonography of Second Trimester, Single Fetus | ICD10PCS | Procedure | ICD10PCS |
| 3037974 | Fetal Narrative [Interpretation] Study observation.general 2nd trimester US | Clinical Observation | Measurement | LOINC |
| 3031374 | Fetal Narrative [Interpretation] Study observation.general 2nd trimester, multiple fetuses US | Clinical Observation | Measurement | LOINC |
| 2884808 | Imaging @ Fetus and Obstetrical @ Ultrasonography @ Second Trimester, Single Fetus | ICD10PCS Hierarchy | Procedure | ICD10PCS |
| 2840219 | Imaging @ Fetus and Obstetrical @ Ultrasonography @ Second Trimester, Single Fetus @ None | ICD10PCS Hierarchy | Procedure | ICD10PCS |
| 2852789 | Imaging @ Fetus and Obstetrical @ Ultrasonography @ Second Trimester, Single Fetus @ None @ None | ICD10PCS Hierarchy | Procedure | ICD10PCS |
| 2211747 | Ultrasound, pregnant uterus, real time with image documentation, fetal and maternal evaluation, first trimester (< 14 weeks 0 days), transabdominal approach; single or first gestation | CPT4 | Procedure | CPT4 |
| 2211748 | Ultrasound, pregnant uterus, real time with image documentation, fetal and maternal evaluation, first trimester (< 14 weeks 0 days), transabdominal approach; each additional gestation (List separately in addition to code for primary procedure) | CPT4 | Procedure | CPT4 |
| 2211753 | Ultrasound, pregnant uterus, real time with image documentation, first trimester fetal nuchal translucency measurement, transabdominal or transvaginal approach; single or first gestation | CPT4 | Procedure | CPT4 |
| 2211754 | Ultrasound, pregnant uterus, real time with image documentation, first trimester fetal nuchal translucency measurement, transabdominal or transvaginal approach; each additional gestation (List separately in addition to code for primary procedure) | CPT4 | Procedure | CPT4 |
| 2793348 | Ultrasonography of First Trimester, Multiple Gestation | ICD10PCS | Procedure | ICD10PCS |
| 21493908 | US for multiple gestation pregnancy in first trimester | Clinical Observation | Measurement | LOINC |
| 2879606 | Imaging @ Fetus and Obstetrical @ Ultrasonography @ First Trimester, Multiple Gestation | ICD10PCS Hierarchy | Procedure | ICD10PCS |
| 2898831 | Imaging @ Fetus and Obstetrical @ Ultrasonography @ First Trimester, Multiple Gestation @ None | ICD10PCS Hierarchy | Procedure | ICD10PCS |
| 21493909 | US transabdominal and transvaginal for multiple gestation pregnancy in first trimester | Clinical Observation | Measurement | LOINC |
| 2840218 | Imaging @ Fetus and Obstetrical @ Ultrasonography @ First Trimester, Multiple Gestation @ None @ None | ICD10PCS Hierarchy | Procedure | ICD10PCS |
| 21494044 | US for pregnancy in first trimester | Clinical Observation | Measurement | LOINC |
| 21494042 | US transabdominal and transvaginal for pregnancy in first trimester | Clinical Observation | Measurement | LOINC |
| 36203523 | Traveled outside the U.S. during first trimester of pregnancy of Mother | Clinical Observation | Measurement | LOINC |
| 3050129 | First trimester maternal screen panel - Serum or Plasma | Lab Test | Measurement | LOINC |
| 3030256 | First trimester maternal screen with nuchal translucency [Interpretation] | Clinical Observation | Observation | LOINC |
| 3031648 | First trimester maternal screen with nuchal translucency [Interpretation] Narrative | Clinical Observation | Observation | LOINC |
| 3050402 | First trimester maternal screen with nuchal translucency panel | Lab Test | Measurement | LOINC |
| 4113139 | First trimester | Qualifier Value | Observation | SNOMED |
| 3657563 | First trimester bleeding | Clinical Finding | Condition | SNOMED |
| 4239938 | First trimester pregnancy | Clinical Finding | Condition | SNOMED |
| 4078393 | Miscarriage in first trimester | Clinical Finding | Condition | SNOMED |
| 40488298 | Ultrasonography in first trimester | Procedure | Procedure | SNOMED |
| 4252252 | Threatened miscarriage in first trimester | Clinical Finding | Condition | SNOMED |
| 4034340 | Prenatal state of fetus, 1st trimester | Clinical Finding | Condition | SNOMED |
| 45763590 | Induced termination of pregnancy in first trimester | Clinical Finding | Condition | SNOMED |
| 42538969 | First trimester Down screening blood test abnormal | Clinical Finding | Condition | SNOMED |
| 40480885 | Education about folic acid in first trimester | Procedure | Procedure | SNOMED |
| 43020954 | Termination of pregnancy after first trimester | Procedure | Procedure | SNOMED |
| 4087135 | Surgical treatment of missed miscarriage of first trimester | Procedure | Procedure | SNOMED |
| 2110326 | Treatment of missed abortion, completed surgically; first trimester | CPT4 | Procedure | CPT4 |
| 36304648 | Cigarettes smoked per day by Mother--1st trimester | Survey | Observation | LOINC |
| 2793347 | Ultrasonography of First Trimester, Single Fetus | ICD10PCS | Procedure | ICD10PCS |
| 3034062 | Fetal Narrative [Interpretation] Study observation.general 1st trimester US | Clinical Observation | Measurement | LOINC |
| 3037993 | Fetal Narrative [Interpretation] Study observation.general transvaginal 1st trimester US | Clinical Observation | Measurement | LOINC |
| 3034647 | Fetal Narrative [Interpretation] Study observation.general 1st trimester, multiple fetuses US | Clinical Observation | Measurement | LOINC |
| 2892608 | Imaging @ Fetus and Obstetrical @ Ultrasonography @ First Trimester, Single Fetus | ICD10PCS Hierarchy | Procedure | ICD10PCS |
| 2892609 | Imaging @ Fetus and Obstetrical @ Ultrasonography @ First Trimester, Single Fetus @ None | ICD10PCS Hierarchy | Procedure | ICD10PCS |
| 2807692 | Imaging @ Fetus and Obstetrical @ Ultrasonography @ First Trimester, Single Fetus @ None @ None | ICD10PCS Hierarchy | Procedure | ICD10PCS |
| 4181468 | Gestation 9- 13 weeks | Clinical Finding | Condition | SNOMED |
| 44791171 | 9 - 13 weeks gestational age | Clinical Finding | Condition | SNOMED |
| 4245908 | Gestation period, 9 weeks | Clinical Finding | Condition | SNOMED |
| 4270513 | Gestation period, 7 weeks | Clinical Finding | Condition | SNOMED |
| 4313026 | Gestation period, 6 weeks | Clinical Finding | Condition | SNOMED |
| 4290009 | Gestation period, 5 weeks | Clinical Finding | Condition | SNOMED |
| 444067 | Gestation period, 42 weeks | Clinical Finding | Condition | SNOMED |
| 442769 | Gestation period, 41 weeks | Clinical Finding | Condition | SNOMED |
| 45773507 | Post-term pregnancy of 40 to 42 weeks | Clinical Finding | Condition | SNOMED |
| 444098 | Gestation period, 40 weeks | Clinical Finding | Condition | SNOMED |
| 435655 | Gestation period, 39 weeks | Clinical Finding | Condition | SNOMED |
| 443871 | Gestation period, 38 weeks | Clinical Finding | Condition | SNOMED |
| 45757118 | Spontaneous onset of labor between 37 and 39 weeks gestation with planned cesarean section | Clinical Finding | Condition | SNOMED |
| 442355 | Gestation period, 37 weeks | Clinical Finding | Condition | SNOMED |
| 40757033 | Group B Streptococcus (GBS) screening documented as performed during week 35-37 gestation (Pre-Cr) | CPT4 | Observation | CPT4 |
| 444267 | Gestation period, 35 weeks | Clinical Finding | Condition | SNOMED |
| 443874 | Gestation period, 34 weeks | Clinical Finding | Condition | SNOMED |
| 441678 | Gestation period, 33 weeks | Clinical Finding | Condition | SNOMED |
| 433864 | Gestation period, 31 weeks | Clinical Finding | Condition | SNOMED |
| 434484 | Gestation period, 30 weeks | Clinical Finding | Condition | SNOMED |
| 4326232 | Gestation period, 3 weeks | Clinical Finding | Condition | SNOMED |
| 444417 | Gestation period, 29 weeks | Clinical Finding | Condition | SNOMED |
| 44817054 | Mother's Non-treponemal or treponemal test was performed at 28-32 weeks gestation [CDC.CS] | Survey | Observation | LOINC |
| 4180111 | Third trimester pregnancy less than 36 weeks | Clinical Finding | Condition | SNOMED |
| 432430 | Gestation period, 27 weeks | Clinical Finding | Condition | SNOMED |
| 444023 | Gestation period, 26 weeks | Clinical Finding | Condition | SNOMED |
| 435640 | Gestation period, 25 weeks | Clinical Finding | Condition | SNOMED |
| 4336226 | Gestation period, 23 weeks | Clinical Finding | Condition | SNOMED |
| 4274955 | Gestation period, 22 weeks | Clinical Finding | Condition | SNOMED |
| 4185780 | Gestation period, 21 weeks | Clinical Finding | Condition | SNOMED |
| 4220085 | Gestation period, 2 weeks | Clinical Finding | Condition | SNOMED |
| 4181751 | Gestation period, 19 weeks | Clinical Finding | Condition | SNOMED |
| 44790206 | Mid trimester scan | Procedure | Procedure | SNOMED |
| 4097608 | Gestation period, 18 weeks | Clinical Finding | Condition | SNOMED |
| 4277749 | Gestation period, 17 weeks | Clinical Finding | Condition | SNOMED |
| 4283690 | Gestation period, 15 weeks | Clinical Finding | Condition | SNOMED |
| 4178165 | Gestation 14 - 20 weeks | Clinical Finding | Condition | SNOMED |
| 44791170 | 14 - 20 weeks gestational age | Clinical Finding | Condition | SNOMED |
| 4248725 | Gestation period, 14 weeks | Clinical Finding | Condition | SNOMED |
| 4266517 | Gestation period, 13 weeks | Clinical Finding | Condition | SNOMED |
| 4174506 | Gestation period, 11 weeks | Clinical Finding | Condition | SNOMED |
| 4242241 | Gestation period, 10 weeks | Clinical Finding | Condition | SNOMED |
| 4337360 | Gestation period, 1 week | Clinical Finding | Condition | SNOMED |
| 762907 | Gestation period greater than or equal to 37 weeks | Clinical Finding | Condition | SNOMED |
| 4062558 | False labor at or after 37 completed weeks of gestation | Clinical Finding | Condition | SNOMED |
| 4322726 | Gestation less than 9 weeks | Clinical Finding | Condition | SNOMED |
| 44791172 | Under 9 weeks gestational age | Clinical Finding | Condition | SNOMED |
| 438543 | Gestation period, 36 weeks | Clinical Finding | Condition | SNOMED |
| 442558 | Gestation period, 32 weeks | Clinical Finding | Condition | SNOMED |
| 444461 | Gestation period, 28 weeks | Clinical Finding | Condition | SNOMED |
| 439922 | Gestation period, 24 weeks | Clinical Finding | Condition | SNOMED |
| 4051642 | Gestation period, 20 weeks | Clinical Finding | Condition | SNOMED |
| 4049621 | Gestation period, 16 weeks | Clinical Finding | Condition | SNOMED |
| 4197245 | Gestation period, 12 weeks | Clinical Finding | Condition | SNOMED |
| 4132434 | Gestation period, 8 weeks | Clinical Finding | Condition | SNOMED |
| 4195157 | Gestation period, 4 weeks | Clinical Finding | Condition | SNOMED |

Appendix 2. ICD, CPT, and DRG codes suggestive of childbirth delivery dates.

| Medical Language System | Code | Description |
|---|---|---|
| ICD-10_DX | Z37.0 | Single live birth |
| ICD-10_DX | Z37.1 | Single stillbirth |
| ICD-10_DX | Z37.2 | Twins, both liveborn |
| ICD-10_DX | Z37.3 | Twins, one liveborn and one stillborn |
| ICD-10_DX | Z37.4 | Twins, both stillborn |
| ICD-10_DX | Z37.5 | Other multiple births, all liveborn |
| ICD-10_DX | Z37.6 | Other multiple births, some liveborn |
| ICD-10_DX | Z37.7 | Other multiple births, all stillborn |
| ICD-10_DX | Z37.9 | Outcome of delivery, unspecified |
| ICD-10_DX | O80 | Encounter for full-term uncomplicated delivery |
| ICD-10_DX | O82 | Encounter for cesarean delivery without indication |
| ICD-10_DX | O7582 | Onset (spontaneous) of labor after 37 completed weeks of gestation but before 39 completed weeks gestation, with delivery by (planned) cesarean section |
| ICD_10_PCS | 10D07Z3 | Extraction of Products of Conception, Low Forceps, Via Natural or Artificial Opening |
| ICD_10_PCS | 10D07Z4 | Extraction of Products of Conception, Mid Forceps, Via Natural or Artificial Opening |
| ICD_10_PCS | 10D07Z5 | Extraction of Products of Conception, High Forceps, Via Natural or Artificial Opening |
| ICD_10_PCS | 10D07Z6 | Extraction of Products of Conception, Vacuum, Via Natural or Artificial Opening |
| ICD_10_PCS | 10D07Z7 | Extraction of Products of Conception, Internal Version, Via Natural or Artificial Opening |
| ICD_10_PCS | 10D07Z8 | Extraction of Products of Conception, Other, Via Natural or Artificial Opening |
| ICD_10_PCS | 10E0XZZ | Delivery of Products of Conception, External Approach |
| ICD_10_PCS | 10D00Z0 | Extraction of Products of Conception, High, Open Approach |
| ICD_10_PCS | 10D00Z1 | Extraction of Products of Conception, Low, Open Approach |
| ICD_10_PCS | 10D00Z2 | Extraction of Products of Conception, Extraperitoneal, Open Approach |
| DRG | 765 | CESAREAN SECTION WITH CC/MCC |
| DRG | 766 | CESAREAN SECTION WITHOUT CC/MCC |
| DRG | 767 | VAGINAL DELIVERY WITH STERILIZATION AND/OR D&C |
| DRG | 768 | VAGINAL DELIVERY WITH O.R. PROCEDURES EXCEPT STERILIZATION AND/OR D&C |
| DRG | 774 | VAGINAL DELIVERY WITH COMPLICATING DIAGNOSES |
| DRG | 775 | VAGINAL DELIVERY WITHOUT COMPLICATING DIAGNOSES |
| DRG | 783 | CESAREAN SECTION WITH STERILIZATION WITH MCC |
| DRG | 784 | CESAREAN SECTION WITH STERILIZATION WITH CC |
| DRG | 785 | CESAREAN SECTION WITH STERILIZATION WITHOUT CC/MCC |
| DRG | 786 | CESAREAN SECTION WITHOUT STERILIZATION WITH MCC |
| DRG | 787 | CESAREAN SECTION WITHOUT STERILIZATION WITH CC |
| DRG | 788 | CESAREAN SECTION WITHOUT STERILIZATION WITHOUT CC/MCC |
| DRG | 796 | VAGINAL DELIVERY WITH STERILIZATION/D&C WITH MCC |
| DRG | 797 | VAGINAL DELIVERY WITH STERILIZATION AND/OR D&C WITH CC |
| DRG | 798 | VAGINAL DELIVERY WITH STERILIZATION/D&C WITHOUT CC/MCC |
| DRG | 805 | VAGINAL DELIVERY WITHOUT STERILIZATION/D&C WITH MCC |
| DRG | 806 | VAGINAL DELIVERY WITHOUT STERILIZATION/D&C WITH CC |
| DRG | 807 | VAGINAL DELIVERY WITHOUT STERILIZATION OR D&C WITHOUT CC/MCC |
| CPT | 59400 | Vaginal Delivery, Antepartum and Postpartum Care Procedures |
| CPT | 59409 | Vaginal delivery only (with or without episiotomy and/or forceps) |
| CPT | 59410 | Vaginal delivery only (with or without episiotomy and/or forceps) |
| CPT | 59514 | Cesarean delivery only |
| CPT | 59610 | Delivery Procedures After Previous Cesarean Delivery |
| CPT | 59612 | Vaginal delivery only, after previous cesarean delivery (with or without episiotomy and/or forceps) |
| CPT | 59614 | Vaginal delivery only, after previous cesarean delivery (with or without episiotomy and/or forceps) |
| CPT | 59620 | Cesarean delivery only, following attempted vaginal delivery after previous cesarean delivery |

Appendix 3. OMOP CDM concepts for delivery date-related EHR records.

| Concept ID | Concept Name | Class | Domain | Vocabulary |
|---|---|---|---|---|
| 443445 | Outcome of delivery - finding | clinical finding | Condition | SNOMED |
| 4145318 | Outcome of delivery | clinical finding | Observation | SNOMED |
| 4163851 | Multiple birth | clinical finding | Condition | SNOMED |
| 4014456 | Triplets - all live born | clinical finding | Condition | SNOMED |
| 45757166 | Quadruplets, all live born | clinical finding | Condition | SNOMED |
| 45757167 | Quintuplets, all live born | clinical finding | Condition | SNOMED |
| 45772082 | Sextuplets, all live born | clinical finding | Condition | SNOMED |
| 4094046 | Triplet birth | clinical finding | Condition | SNOMED |
| 45765500 | Quadruplet birth | clinical finding | Condition | SNOMED |
| 45765501 | Quintuplet birth | clinical finding | Condition | SNOMED |
| 45765502 | Sextuplet birth | clinical finding | Condition | SNOMED |
| 4014295 | Single live birth | clinical finding | Condition | SNOMED |
| 4014454 | Single stillbirth | clinical finding | Condition | SNOMED |
| 4014296 | Twins - both live born | clinical finding | Condition | SNOMED |
| 4014455 | Twins - one still and one live born | clinical finding | Condition | SNOMED |
| 4015162 | Twins - both stillborn | clinical finding | Condition | SNOMED |
| 441641 | Delivery normal | clinical finding | Condition | SNOMED |
| 4205240 | Spontaneous vertex delivery | clinical finding | Condition | SNOMED |
| 4073422 | Spontaneous breech delivery | procedure | Procedure | SNOMED |
| 193277 | Deliveries by cesarean | clinical finding | Condition | SNOMED |
| 4061457 | Delivery by elective cesarean section | clinical finding | Condition | SNOMED |
| 4066112 | Delivery by emergency cesarean section | clinical finding | Condition | SNOMED |
| 4061458 | Delivery by cesarean hysterectomy | clinical finding | Condition | SNOMED |
| 45757118 | Spontaneous onset of labor between 37 and 39 weeks gestation with planned cesarean section | clinical finding | Condition | SNOMED |
| 2784567 | Extraction of Products of Conception, Low Forceps, Via Natural or Artificial Opening | ICD10PCS | Procedure | ICD10PCS |
| 2784568 | Extraction of Products of Conception, Mid Forceps, Via Natural or Artificial Opening | ICD10PCS | Procedure | ICD10PCS |
| 2784569 | Extraction of Products of Conception, High Forceps, Via Natural or Artificial Opening | ICD10PCS | Procedure | ICD10PCS |
| 2784570 | Extraction of Products of Conception, Vacuum, Via Natural or Artificial Opening | ICD10PCS | Procedure | ICD10PCS |
| 2784571 | Extraction of Products of Conception, Internal Version, Via Natural or Artificial Opening | ICD10PCS | Procedure | ICD10PCS |
| 2784572 | Extraction of Products of Conception, Other, Via Natural or Artificial Opening | ICD10PCS | Procedure | ICD10PCS |
| 2784578 | Delivery of Products of Conception, External Approach | ICD10PCS | Procedure | ICD10PCS |
| 2784564 | Extraction of Products of Conception, High, Open Approach | ICD10PCS | Procedure | ICD10PCS |
| 2784565 | Extraction of Products of Conception, Low, Open Approach | ICD10PCS | Procedure | ICD10PCS |
| 2784566 | Extraction of Products of Conception, Extraperitoneal, Open Approach | ICD10PCS | Procedure | ICD10PCS |
| 38001485 | Cesarean section w CC/MCC | MS-DRG (Domain Observation) | Observation | DRG |
| 38001486 | Cesarean section w/o CC/MCC | MS-DRG (Domain Observation) | Observation | DRG |
| 38001487 | Vaginal delivery w sterilization &/or D&C | MS-DRG (Domain Observation) | Observation | DRG |
| 38001488 | Vaginal delivery w O.R. proc except steril &/or D&C | MS-DRG (Domain Observation) | Observation | DRG |
| 38001491 | Vaginal delivery w complicating diagnoses | MS-DRG (Domain Observation) | Observation | DRG |
| 38001492 | Vaginal delivery w/o complicating diagnoses | MS-DRG (Domain Observation) | Observation | DRG |
| 2110307 | Routine obstetric care including antepartum care, vaginal delivery (with or without episiotomy, and/or forceps) and postpartum care | CPT4(Domain Procedure) | Procedure | CPT4 |
| 2110308 | Vaginal delivery only (with or without episiotomy and/or forceps) | CPT4(Domain Procedure) | Procedure | CPT4 |
| 2110309 | Vaginal delivery only (with or without episiotomy and/or forceps); including postpartum care | CPT4(Domain Procedure) | Procedure | CPT4 |
| 2110319 | Routine obstetric care including antepartum care, vaginal delivery (with or without episiotomy, and/or forceps) and postpartum care, after previous cesarean delivery | CPT4(Domain Procedure) | Procedure | CPT4 |
| 2110320 | Vaginal delivery only, after previous cesarean delivery (with or without episiotomy and/or forceps) | CPT4(Domain Procedure) | Procedure | CPT4 |
| 2110321 | Vaginal delivery only, after previous cesarean delivery (with or without episiotomy and/or forceps); including postpartum care | CPT4(Domain Procedure) | Procedure | CPT4 |
| 2110316 | Cesarean delivery only | CPT4(Domain Procedure) | Procedure | CPT4 |
| 2110323 | Cesarean delivery only, following attempted vaginal delivery after previous cesarean delivery | CPT4(Domain Procedure) | Procedure | CPT4 |
| 4264823 | Birth outcome (observable entity) | Observable Entity | Observation | SNOMED |
| 4015270 | Birth of child (finding) | clinical finding | condition | SNOMED |
| 44793347 | Total number of registerable births at delivery | Observable Entity | observation | SNOMED |
| 4063163 | Multiple delivery, all by cesarean section | clinical finding | condition | SNOMED |
| 4063162 | Multiple delivery, all by forceps and vacuum extractor | clinical finding | condition | SNOMED |
| 4059751 | Multiple delivery, all spontaneous | clinical finding | condition | SNOMED |
| 4069200 | Premature birth of multiple newborns | clinical finding | condition | SNOMED |
| 4066292 | Term birth of multiple newborns | clinical finding | condition | SNOMED |
| 4101844 | Twin birth | clinical finding | condition | SNOMED |
| 40482735 | Liveborn born in hospital (situation) | context-dependent | observation | SNOMED |
| 40483126 | Liveborn born in hospital by cesarean section (situation) | context-dependent | observation | SNOMED |
| 42539267 | Multiple liveborn in hospital by vaginal delivery (situation) | context-dependent | observation | SNOMED |
| 36713468 | Multiple liveborn other than twins born in hospital (situation) | context-dependent | observation | SNOMED |
| 40483521 | Single liveborn born in hospital by cesarean section (situation) | context-dependent | observation | SNOMED |
| 36713074 | Single liveborn born in hospital by vaginal delivery (situation) | context-dependent | observation | SNOMED |
| 36713465 | Singleton liveborn born in hospital (situation) | context-dependent | observation | SNOMED |
| 42539210 | Triplet liveborn in hospital by cesarean section (situation) | context-dependent | observation | SNOMED |
| 40483084 | Twin liveborn born in hospital (situation) | context-dependent | observation | SNOMED |
| 40483101 | Twin liveborn born in hospital by cesarean section (situation) | context-dependent | observation | SNOMED |
| 4014719 | Labor details (finding) | drug product | Drug | VANDF |
| 4262313 | Time of delivery (observable entity) | Observable Entity | observation | SNOMED |
| 4014291 | Birth detail (observable entity) | drug product | Drug | VANDF |
| 4250009 | Born by breech delivery | context-dependent | observation | SNOMED |
| 4192676 | Born by cesarean section | context-dependent | observation | SNOMED |
| 4212794 | Born by elective cesarean section | context-dependent | observation | SNOMED |
| 4250010 | Born by emergency cesarean section | context-dependent | observation | SNOMED |
| 4217586 | Born by forceps delivery | context-dependent | observation | SNOMED |
| 37310369 | Born by high forceps delivery | context-dependent | observation | SNOMED |
| 37310393 | Born by low forceps delivery | context-dependent | observation | SNOMED |
| 37310404 | Born by mid-cavity forceps delivery | context-dependent | observation | SNOMED |
| 4216797 | Born by normal vaginal delivery | context-dependent | observation | SNOMED |
| 4236293 | Born by ventouse delivery | context-dependent | observation | SNOMED |
| 44813089 | Midwife in attendance at birth | context-dependent | observation | SNOMED |
| 44802498 | Midwife not in attendance at birth | context-dependent | observation | SNOMED |
| 4088584 | Caul membrane over baby's head at delivery | context-dependent | observation | SNOMED |
| 4128845 | Delivery problem | clinical finding | Condition | SNOMED |
| 4200201 | Finding of birth outcome | clinical finding | condition | SNOMED |
| 4126390 | Finding of pattern of delivery | clinical finding | condition | SNOMED |
| 4122720 | Finding of second stage of labor | clinical finding | condition | SNOMED |
| 4096383 | Finding of speed of delivery | clinical finding | condition | SNOMED |
| 4125602 | Finding of third stage of labor | clinical finding | condition | SNOMED |
| 4009879 | Good neonatal condition at birth | clinical finding | condition | SNOMED |
| 433260 | Mother delivered | clinical finding | condition | SNOMED |
| 3174660 | Traumatic birth | clinical finding | condition | Nebraska Lexicon |
| 4014720 | Normal birth | clinical finding | condition | SNOMED |
| 4118903 | Normal delivery - occipitoanterior | clinical finding | condition | SNOMED |
| 4063160 | Normal delivery but ante- or post-natal conditions present | clinical finding | condition | SNOMED |
| 442069 | Vacuum extractor delivery - delivered | clinical finding | condition | SNOMED |
| 4065737 | Delivery by combination of forceps and vacuum extractor | clinical finding | condition | SNOMED |
| 45757174 | Vacuum assisted vaginal delivery | clinical finding | condition | SNOMED |
| 4216316 | Birth | clinical finding | condition | SNOMED |
| 4009589 | Post-term delivery | clinical finding | condition | SNOMED |
| 4086393 | Premature delivery | clinical finding | condition | SNOMED |
| 4092289 | Livebirth | clinical finding | condition | SNOMED |
| 4272248 | Premature birth of newborn | clinical finding | condition | SNOMED |
| 443213 | Stillbirth | clinical finding | condition | SNOMED |
| 4054968 | Term birth of newborn | clinical finding | condition | SNOMED |
| 443445 | Outcome of delivery - finding | clinical finding | Condition | SNOMED |
| 4145318 | Outcome of delivery | clinical finding | Observation | SNOMED |
| 4163851 | Multiple birth | clinical finding | Condition | SNOMED |
| 4014456 | Triplets - all live born | clinical finding | Condition | SNOMED |
| 45757166 | Quadruplets, all live born | clinical finding | Condition | SNOMED |
| 45757167 | Quintuplets, all live born | clinical finding | Condition | SNOMED |
| 45772082 | Sextuplets, all live born | clinical finding | Condition | SNOMED |
| 4094046 | Triplet birth | clinical finding | Condition | SNOMED |
| 45765500 | Quadruplet birth | clinical finding | Condition | SNOMED |
| 45765501 | Quintuplet birth | clinical finding | Condition | SNOMED |
| 45765502 | Sextuplet birth | clinical finding | Condition | SNOMED |
| 4014295 | Single live birth | clinical finding | Condition | SNOMED |
| 4014454 | Single stillbirth | clinical finding | Condition | SNOMED |
| 4014296 | Twins - both live born | clinical finding | Condition | SNOMED |
| 4014455 | Twins - one still and one live born | clinical finding | Condition | SNOMED |
| 4015162 | Twins - both stillborn | clinical finding | Condition | SNOMED |
| 441641 | Delivery normal | clinical finding | Condition | SNOMED |
| 4205240 | Spontaneous vertex delivery | clinical finding | Condition | SNOMED |
| 4073422 | Spontaneous breech delivery | procedure | Procedure | SNOMED |
| 193277 | Deliveries by cesarean | clinical finding | Condition | SNOMED |
| 4061457 | Delivery by elective cesarean section | clinical finding | Condition | SNOMED |
| 4066112 | Delivery by emergency cesarean section | clinical finding | Condition | SNOMED |
| 4061458 | Delivery by cesarean hysterectomy | clinical finding | Condition | SNOMED |
| 45757118 | Spontaneous onset of labor between 37 and 39 weeks gestation with planned cesarean section | clinical finding | Condition | SNOMED |
| 2784567 | Extraction of Products of Conception, Low Forceps, Via Natural or Artificial Opening | ICD10PCS | Procedure | ICD10PCS |
| 2784568 | Extraction of Products of Conception, Mid Forceps, Via Natural or Artificial Opening | ICD10PCS | Procedure | ICD10PCS |
| 2784569 | Extraction of Products of Conception, High Forceps, Via Natural or Artificial Opening | ICD10PCS | Procedure | ICD10PCS |
| 2784570 | Extraction of Products of Conception, Vacuum, Via Natural or Artificial Opening | ICD10PCS | Procedure | ICD10PCS |
| 2784571 | Extraction of Products of Conception, Internal Version, Via Natural or Artificial Opening | ICD10PCS | Procedure | ICD10PCS |
| 2784572 | Extraction of Products of Conception, Other, Via Natural or Artificial Opening | ICD10PCS | Procedure | ICD10PCS |
| 2784578 | Delivery of Products of Conception, External Approach | ICD10PCS | Procedure | ICD10PCS |
| 2784564 | Extraction of Products of Conception, High, Open Approach | ICD10PCS | Procedure | ICD10PCS |
| 2784565 | Extraction of Products of Conception, Low, Open Approach | ICD10PCS | Procedure | ICD10PCS |
| 2784566 | Extraction of Products of Conception, Extraperitoneal, Open Approach | ICD10PCS | Procedure | ICD10PCS |
| 38001485 | Cesarean section w CC/MCC | MS-DRG (Domain Observation) | Observation | DRG |
| 38001486 | Cesarean section w/o CC/MCC | MS-DRG (Domain Observation) | Observation | DRG |
| 38001487 | Vaginal delivery w sterilization &/or D&C | MS-DRG (Domain Observation) | Observation | DRG |
| 38001488 | Vaginal delivery w O.R. proc except steril &/or D&C | MS-DRG (Domain Observation) | Observation | DRG |
| 38001491 | Vaginal delivery w complicating diagnoses | MS-DRG (Domain Observation) | Observation | DRG |
| 38001492 | Vaginal delivery w/o complicating diagnoses | MS-DRG (Domain Observation) | Observation | DRG |
| 2110307 | Routine obstetric care including antepartum care, vaginal delivery (with or without episiotomy, and/or forceps) and postpartum care | CPT4(Domain Procedure) | Procedure | CPT4 |
| 2110308 | Vaginal delivery only (with or without episiotomy and/or forceps) | CPT4(Domain Procedure) | Procedure | CPT4 |
| 2110309 | Vaginal delivery only (with or without episiotomy and/or forceps); including postpartum care | CPT4(Domain Procedure) | Procedure | CPT4 |
| 2110319 | Routine obstetric care including antepartum care, vaginal delivery (with or without episiotomy, and/or forceps) and postpartum care, after previous cesarean delivery | CPT4(Domain Procedure) | Procedure | CPT4 |
| 2110320 | Vaginal delivery only, after previous cesarean delivery (with or without episiotomy and/or forceps) | CPT4(Domain Procedure) | Procedure | CPT4 |
| 2110321 | Vaginal delivery only, after previous cesarean delivery (with or without episiotomy and/or forceps); including postpartum care | CPT4(Domain Procedure) | Procedure | CPT4 |
| 2110316 | Cesarean delivery only | CPT4(Domain Procedure) | Procedure | CPT4 |
| 2110323 | Cesarean delivery only, following attempted vaginal delivery after previous cesarean delivery | CPT4(Domain Procedure) | Procedure | CPT4 |
| 4264823 | Birth outcome (observable entity) | Observable Entity | Observation | SNOMED |
| 4015270 | Birth of child (finding) | clinical finding | condition | SNOMED |
| 44793347 | Total number of registerable births at delivery | Observable Entity | observation | SNOMED |
| 4063163 | Multiple delivery, all by cesarean section | clinical finding | condition | SNOMED |
| 4063162 | Multiple delivery, all by forceps and vacuum extractor | clinical finding | condition | SNOMED |
| 4059751 | Multiple delivery, all spontaneous | clinical finding | condition | SNOMED |
| 4069200 | Premature birth of multiple newborns | clinical finding | condition | SNOMED |
| 4066292 | Term birth of multiple newborns | clinical finding | condition | SNOMED |
| 4101844 | Twin birth | clinical finding | condition | SNOMED |
| 40482735 | Liveborn born in hospital (situation) | context-dependent | observation | SNOMED |
| 40483126 | Liveborn born in hospital by cesarean section (situation) | context-dependent | observation | SNOMED |
| 42539267 | Multiple liveborn in hospital by vaginal delivery (situation) | context-dependent | observation | SNOMED |
| 36713468 | Multiple liveborn other than twins born in hospital (situation) | context-dependent | observation | SNOMED |
| 40483521 | Single liveborn born in hospital by cesarean section (situation) | context-dependent | observation | SNOMED |
| 36713074 | Single liveborn born in hospital by vaginal delivery (situation) | context-dependent | observation | SNOMED |
| 36713465 | Singleton liveborn born in hospital (situation) | context-dependent | observation | SNOMED |
| 42539210 | Triplet liveborn in hospital by cesarean section (situation) | context-dependent | observation | SNOMED |
| 40483084 | Twin liveborn born in hospital (situation) | context-dependent | observation | SNOMED |
| 40483101 | Twin liveborn born in hospital by cesarean section (situation) | context-dependent | observation | SNOMED |
| 4014719 | Labor details (finding) | drug product | Drug | VANDF |
| 4262313 | Time of delivery (observable entity) | Observable Entity | observation | SNOMED |
| 4014291 | Birth detail (observable entity) | drug product | Drug | VANDF |
| 4250009 | Born by breech delivery | context-dependent | observation | SNOMED |
| 4192676 | Born by cesarean section | context-dependent | observation | SNOMED |
| 4212794 | Born by elective cesarean section | context-dependent | observation | SNOMED |
| 4250010 | Born by emergency cesarean section | context-dependent | observation | SNOMED |
| 4217586 | Born by forceps delivery | context-dependent | observation | SNOMED |
| 37310369 | Born by high forceps delivery | context-dependent | observation | SNOMED |
| 37310393 | Born by low forceps delivery | context-dependent | observation | SNOMED |
| 37310404 | Born by mid-cavity forceps delivery | context-dependent | observation | SNOMED |
| 4216797 | Born by normal vaginal delivery | context-dependent | observation | SNOMED |
| 4236293 | Born by ventouse delivery | context-dependent | observation | SNOMED |
| 44813089 | Midwife in attendance at birth | context-dependent | observation | SNOMED |
| 44802498 | Midwife not in attendance at birth | context-dependent | observation | SNOMED |
| 4088584 | Caul membrane over baby's head at delivery | context-dependent | observation | SNOMED |
| 4128845 | Delivery problem | clinical finding | Condition | SNOMED |
| 4200201 | Finding of birth outcome | clinical finding | condition | SNOMED |
| 4126390 | Finding of pattern of delivery | clinical finding | condition | SNOMED |
| 4122720 | Finding of second stage of labor | clinical finding | condition | SNOMED |
| 4096383 | Finding of speed of delivery | clinical finding | condition | SNOMED |
| 4125602 | Finding of third stage of labor | clinical finding | condition | SNOMED |
| 4009879 | Good neonatal condition at birth | clinical finding | condition | SNOMED |
| 433260 | Mother delivered | clinical finding | condition | SNOMED |
| 3174660 | Traumatic birth | clinical finding | condition | Nebraska Lexicon |
| 4014720 | Normal birth | clinical finding | condition | SNOMED |
| 4118903 | Normal delivery - occipitoanterior | clinical finding | condition | SNOMED |
| 4063160 | Normal delivery but ante- or post-natal conditions present | clinical finding | condition | SNOMED |
| 442069 | Vacuum extractor delivery - delivered | clinical finding | condition | SNOMED |
| 4065737 | Delivery by combination of forceps and vacuum extractor | clinical finding | condition | SNOMED |
| 45757174 | Vacuum assisted vaginal delivery | clinical finding | condition | SNOMED |
| 4216316 | Birth | clinical finding | condition | SNOMED |
| 4009589 | Post-term delivery | clinical finding | condition | SNOMED |
| 4086393 | Premature delivery | clinical finding | condition | SNOMED |
| 4092289 | Livebirth | clinical finding | condition | SNOMED |
| 4272248 | Premature birth of newborn | clinical finding | condition | SNOMED |
| 443213 | Stillbirth | clinical finding | condition | SNOMED |
| 4054968 | Term birth of newborn | clinical finding | condition | SNOMED |